\documentstyle[12pt,aaspp4,epsfig,flushrt]{article}  
\newcommand{\sig}{\:\lower0.6ex\hbox{$\stackrel{\textstyle >}{\sim}$}\:}
\newcommand{\sil}{\:\lower0.6ex\hbox{$\stackrel{\textstyle <}{\sim}$}\:}

\begin{document}

\title{Gravitational Collapse in Turbulent Molecular
Clouds. I. Gasdynamical Turbulence.}

\author{Ralf S. Klessen\altaffilmark{1,2}}
\author{Fabian Heitsch\altaffilmark{2}}
\author{Mordecai-Mark Mac Low\altaffilmark{2,3}}
\altaffiltext{1}{Sterrewacht Leiden, Postbus 9613, 2300-RA Leiden, The
Netherlands; E-mail: klessen@strw.leidenuniv.nl} 
\altaffiltext{2}{Max-Planck-Institut f\"ur Astronomie, K\"onigstuhl 17, D-69117
Heidelberg, Germany; E-mail: heitsch@mpia-hd.mpg.de}
\altaffiltext{3}{Department of Astrophysics, American Museum of
Natural History, Central Park West at 79th Street, New York, New York
10024-5192, USA; E-mail: mordecai@amnh.org}

\author{(accepted to {\em The Astrophysical Journal})}

\begin{abstract}
Observed molecular clouds often appear to have very low star formation
efficiencies and lifetimes an order of magnitude longer than their
free-fall times.  Their support is attributed to the random supersonic
motions observed in them.  We study the support of molecular clouds
against gravitational collapse by supersonic, gas dynamical turbulence
using direct numerical simulation.  Computations with two different
algorithms are compared: a particle-based, Lagrangian method (SPH),
and a grid-based, Eulerian, second-order method (ZEUS).  The effects
of both algorithm and resolution can be studied with this method.  We
find that, under typical molecular cloud conditions, global collapse
can indeed be prevented, but density enhancements caused by strong
shocks nevertheless become gravitationally unstable and collapse into
dense cores and, presumably, stars.  The occurance and efficiency of
local collapse decreases as the driving wave length decreases and the
driving strength increases.  It appears that local collapse can only
be prevented entirely with unrealistically short wave length driving,
but observed core formation rates can be reproduced with more
realistic driving.  At high collapse rates, cores are formed on short
time scales in coherent structures with high efficiency, while at low
collapse rates they are scattered randomly throughout the region and
exhibit considerable age spread.  We suggest that this naturally
explains the observed distinction between isolated and clustered star
formation.
\end{abstract}


\clearpage
\section{Motivation}

All presently known star formation occurs in cold molecular clouds.
Application of the pioneering work of Jeans (1902) on the stability of
self-gravitating gaseous systems shows that observed molecular clouds
vastly exceed the critical mass for gravitational collapse.  Thus,
clouds should efficiently form stars on a free-fall time scale of the
order of $\tau_{\rm ff} \sim 10^6$ years in the absence of other
effects.  However, the lifetime of a typical molecular cloud is
generally believed to be a factor of ten or twenty longer than
predicted by Jeans' classical theory (Blitz \& Shu 1980), although we
note that this is subject to controversy.  Ballesteros-Paredes,
Hartmann, \& V\'azquez-Semadeni (1999) and Elmegreen (2000) for
example have argued that not only is the internal structure of
molecular clouds transient, but that the clouds as a whole may be
rather short-lived objects.  They suggested that lifetimes of order of
$\tau_{\rm ff}$ may actually be necessary to explain the lack of
10 -- 20 million year old T Tauri stars associated with some molecular
clouds.  It is observed that stars often do not form in one
`catastrophic' event associated with the global collapse of the entire
cloud.  Instead, they form in very localized regions dispersed through
an {\em apparently} stable cloud (for an overview see Williams, Blitz
\& McKee 1999 and references therein). The total efficiency of
conversion from gas into stars in typical molecular clouds is very
low---of the order of a few percent (e.g.\ Duerr, Imhoff, \& Lada
1982; Leisawitz, Bash, \& Thaddeus 1989).  A comprehensive
astrophysical explanation remains elusive and its discovery remains
one of the great challenges for any theory of star formation. 

Molecular clouds are  turbulent. This is an essential ingredient
for understanding their properties and characteristic spatial and
temporal behavior.  Turbulent gas motions are highly supersonic as
indicated by the superthermal line widths  observed
throughout molecular clouds (Williams et al.~1999). The kinetic energy
carried in that motion is sufficient to balance the potential energy
of the cloud, presumably halting global collapse, a proposition that
we will test in this paper.  However, it can be shown that
interstellar turbulence decays quite rapidly, on time scales of the
order of the free-fall time of the system 
\begin{equation}
\label{eqn:free-fall-time}
\tau_{\rm ff} = \sqrt{\frac{3 \pi}{32 G \rho}}\;
\end{equation}
(Mac~Low et al.~1998, Stone, Ostriker \& Gammie 1998, Mac Low 1999, see also Porter,
Pouquet \& Woodward 1992a,b, 1994 and Padoan \& Nordlund 1999).  Strictly
speaking, equation \ref{eqn:free-fall-time} is only valid for
spherical perturbations with homogeneous density $\rho$, with $G$
denoting the gravitational constant. However, for more general
geometries or density distributions equation \ref{eqn:free-fall-time}
still gives a good approximation to $\tau_{\rm ff}$ if we take
$\rho$ to be the {\em mean} density of the system.  To explain their
observed long life times, turbulence in molecular clouds must be
constantly driven (Gammie \& Ostriker 1996, Mac~Low 1999). The
interplay between self-gravity on the one hand (leading to
local collapse and star formation) and turbulent gas motion on the
other hand (trying to prevent this process) appears to play a key role
in regulating the structure of molecular clouds and determining where
and when stars form.

Unfortunately, a theory of compressible turbulence complete enough to
fully address the issue of stability against gravitational collapse
does not exist, nor is it even visible on the horizon.  A variety of
schemes have been proposed to incorporate the effect of incompressible
turbulence into a gravitational stability analysis.  However,
molecular clouds are extremely compressible. Moreover, the
approximations necessary for solution of the resulting equations are
very stringent and appear to severely limit their applicability to the
physical conditions found in interstellar clouds.  This situation
demands a thorough numerical approach.  Although still far away from
fully describing all phenomena present in molecular clouds, numerical
modeling can capture the important features of supersonic,
compressible turbulence in self-gravitating, ideal gases.  In this
paper, we perform a numerical Jeans analysis for self-gravitating,
compressible, turbulent gas, and apply the result to molecular clouds
and star forming regions.  We do not include magnetic fields here, but
the work presented here provides the foundation for studies including
magnetic fields. Preliminary results appear not to reach markedly
different conclusions (Mac Low, Klessen, \& Heitsch 1999).

In \S~\ref{sec:jeans} we summarize previous work on the question of
stability against collapse in self-gravitating turbulent media.  Then
in \S~\ref{sec:method} we describe our numerical schemes and
models. The dynamical evolution of our models is discussed in
\S~\ref{sec:collapse}, which also introduces the concept of local
versus global collapse. In \S~\ref{sec:wave-mode-analysis} we perform
a Fourier analysis to quantify the collapse behavior on different
spatial scales. Section \ref{sec:SF-in-MC} explores the implications
of our results for star formation in molecular clouds.  We speculate
about the difference between the `clustered' and `isolated' modes of
star formation and about the different time scales involved.  Finally
in \S~\ref{sec:conclusions} we summarize our work.

\section{Jeans Analysis}
\label{sec:jeans}

A first statement about cloud stability can be made from 
considering the virial theorem. Na{\"\i}vely speaking, in equilibrium
the total kinetic energy in the system adds up to half its potential
energy, $E_{\rm kin} + 1/2\,E_{\rm pot} = 0$. If $E_{\rm kin} +
1/2\,E_{\rm pot}<0$ the system collapses, while $E_{\rm kin} +
1/2\,E_{\rm pot}>0$ implies expansion. In turbulent clouds, the total
kinetic energy includes not only  the internal energy but also the
contributions from turbulent gas motions. If this is taken into
account, simple energy considerations can already provide a qualitative
description of the collapse behavior of turbulent self-gravitating
media (Bonazzola et al.~1987).

A more thorough investigation, however, requires a linear stability
analysis.  For the case of an isothermal, infinite, homogeneous,
self-gravitating medium at rest (i.e.~without turbulent motions) Jeans
(1902) derived a relation between the oscillation frequency $\omega$
and the wave number $k$ of small perturbations,
\begin{equation}
\omega^2 - c_{\rm s}^2 k^2 + 4\pi G\,\rho_0 = 0\;,
\label{eqn:jeans-dispersion-rel}
\end{equation}
where $c_{\rm s}$ is the isothermal sound speed, $G$ the gravitational
constant, and $\rho_0$ the initial mass density. Note that the
derivation includes the ad hoc assumption that the linearized version
of the Poisson equation describes only the relation between the
perturbed potential and the perturbed density, neglecting the potential of
the homogeneous solution.  This is the so-called `Jeans
swindle'.  The third term in equation~\ref{eqn:jeans-dispersion-rel} is
responsible for the existence of decaying and growing modes, as pure
sound waves stem from the dispersion relation $\omega^2 - c_{\rm s}^2
k^2 =0$. Perturbations are unstable against gravitational contraction
if their wave number is below a critical value, the Jeans wave number
$k_{\rm J}$, i.e.~if
\begin{equation}
k^2 < k_{\rm J}^2 \equiv \frac{4 \pi G \rho_0}{c_{\rm s}^2}\;, 
\label{eqn:jeans-wave-number}
\end{equation}
or equivalently if the wave length of the perturbation exceeds a
critical size given by $\lambda_{\rm J} \equiv 2 \pi k_{\rm J}^{-1}$.
This directly translates into a mass limit. All perturbations with
masses exceeding the Jeans mass,
\begin{equation}
M_{\rm J} \equiv  \rho_0 \lambda^3 =  \left( \frac{\pi}{G}
\right)^{3/2} \rho_0^{-1/2} {c_{\rm s}^3} ,
\label{eqn:jeans-mass}
\end{equation}
will collapse under their own weight. As we describe the dynamical
evolution of cubic subregions inside molecular clouds, we use the
cubic definition of the Jeans mass. The critical mass for spherical
perturbations is lower by a factor of $\pi/6$. 

Attempts to include the effect of turbulent motions into this analysis were
already being made in the middle of the century by von Weizs\"acker (1943,
1951), who also considered the production of interstellar clouds from the
shocks and density fluctuations in compressible turbulence. A more
quantitative theory was proposed by Chandrasekhar (1951), who studied the
effect of microturbulence on gravitational collapse, assuming that collapse
occurs on scales much greater than the outer scale of turbulence.  He
derived a dispersion relation similar to
equation~\ref{eqn:jeans-dispersion-rel} replacing $c_{\rm s}^2 \longrightarrow
c_{\rm s}^2 + 1/3 \, \langle v^2 \rangle$, where $\langle v^2 \rangle$ is the
overall velocity dispersion due to turbulent motions.  Developments through
the mid-eighties are reviewed by Scalo (1985).  Particularly noteworthy is the
work of Sasao (1973), who may have been the first to quantitatively show that
the generation of density enhancements by turbulence, which Chandrasekhar
(1951) neglected, might be as important as turbulent support.  In a more
recent analysis, Bonazzola et al.~(1987) suggested a wave length dependent
effective sound speed $c_{\rm s}^2(k) = c_{\rm s}^2 + 1/3 \, v^2(k)$, leading
to a dispersion relation
\begin{equation}
\omega^2 - \left (c_{\rm s}^2 + \frac{1}{3}v^2(k)\right) k^2 + 4\pi G\,\rho_0 = 0\;.
\label{eqn:bonazzola-dispersion-rel}
\end{equation}
In this description, the stability of the system depends not only on
the total amount of energy, but also on the wave length distribution of
the energy, since $v^2(k)$ depends on the turbulent power spectrum
$\epsilon(k)$ as
\begin{equation}
v^2(k) \equiv \int_k^\infty \epsilon(k')dk'\;.
\label{eqn:def-v^2}
\end{equation} 
Thus, the system can be stable at some wave lengths, but not at others.
This approach was also adopted by V{\'a}zquez-Semadeni \& Gazol
(1995), who added Larson's (1981) empirical scaling relations to the analysis.

The most elaborate investigation of the stability of turbulent,
self-gravitating gas was made by Bonazzola et al.~(1992), who used
renormalization group theory to derive a dispersion relation with a
generalized, wave number-dependent, effective sound speed and an
effective kinetic viscosity that together account for turbulence at
all wave lengths shorter than the one in question.  They found a
general dispersion relation (their equation~4.13) that, if applied to
turbulence with a power-law energy spectrum $\epsilon(k) = A
k^{-\alpha}$, predicts a critical value of the power-law exponent
$\alpha = 3$.  According to their analysis, turbulence with a spectrum
steeper than this can support a region against collapse at large
scales, and below the thermal Jeans scale, but not in between.  On the
other hand, they claim that turbulence with a shallower slope, as is
expected for incompressible turbulence (Kolmogorov 1941), Burgers
turbulence (Lesieur 1997), or shock dominated flows (Passot, Pouquet
\& Woodward 1988), cannot support clouds against collapse at scales
larger than the thermal Jeans wave length.

These analytical approaches make a strong assumption that
substantially limits their reliability, namely that the equilibrium
state is homogeneous, with constant density $\rho_0$. However,
observations clearly show that molecular clouds are extremely
non-uniform.  In fact, it may even be possible to describe the
equilibrium state as an inherently inhomogeneous thermodynamic
critical point (de~Vega, S{\'a}nchez \& Combes 1996a,b; de~Vega \&
S{\'a}nchez 1999).  As a consequence of the assumption of homogeneity,
the further assumption of microturbulence must then be made. The
largest turbulent scale is significantly smaller than the scale of the
analysis.  Interstellar turbulence, however, does not appear to
exhibit such a cut-off in the power spectrum, but rather extends over
{\em all} spatial scales present in the system.  A further corollary
of the assumption of homogeneity is that the turbulent dynamical
time scale is much shorter than the collapse time scale $\tau_{\rm ff}$,
which is only justified if the assumption of microturbulence holds.

One way to achieve progress and circumvent the restrictions of a
purely analytical approach is to perform numerical simulations.
Bonazzola et al.~(1987), for example, used low resolution ($32 \times
32$ collocation points) calculations with a 2-dimensional spectral
code to support their analytical results.  Also restricted to two
dimensions were the hydrodynamical studies by Passot et al.~(1988),
L{\'e}orat, Passot \& Pouquet (1990), V{\'a}zquez-Semadeni et
al.~(1995) and Ballesteros-Paredes, V{\'a}zquez-Semadeni \& Scalo
(1999), although performed with far higher resolution.  Magnetic
fields were introduced in two dimensions by Passot,
V\'azquez-Semadeni, \& Pouquet (1995), and extended to three
dimensions with self-gravity (though at only $64^3$ resolution) by
V\'azquez-Semadeni, Passot, \& Pouquet (1996).  A careful analysis of
1-dimensional computations including both MHD and self-gravity was
presented by Gammie \& Ostriker (1996), who extended their work to
2.5 dimensions more recently (Ostriker, Gammie, \& Stone 1999).  Preliminary
results of high-resolution ($256^3$ zone) simulations with MHD and
self-gravity have been presented by Mac Low et al. (1999) and by
Ostriker, Gammie, \& Stone (1998).  In the present paper we use
two numerical algorithms to examine the stability properties of
three-dimensional hydrodynamical turbulence at higher resolution than
before.  In subsequent work in preparation we will include magnetic
fields as well.

\section{Numerical Methods}
\label{sec:method}

Direct numerical simulation of the Euler equations for gas flow does
not reach the enormous Reynolds numbers typical of molecular clouds
due to the intrinsic or numerical viscosity of any finite-difference
or spectral method.  However, if the details of behavior at the
dissipation scale do not affect the behavior of larger scales, then
all that is required is a low enough viscosity to separate the two
scales.  Incompressible turbulence appears to behave like this
(e.g. Lesieur 1997).  Resolution studies of energy decay in supersonic
compressible turbulence suggest that it may also be true in this case
(Mac Low et al. 1998a).  The resolution studies we do here also
address this question, as increasing the resolution decreases the
dissipation scale, which is always close to the zone size.

We use both Lagrangian and Eulerian numerical methods to solve the
equations of self-gravitating hydrodynamics in three dimensions (3D)
in an attempt to bracket reality by taking advantage of the strengths
of each approach.  This also gives us some protection against
interpreting numerical artifacts as physical effects.  The Lagrangian
method we use is smoothed particle hydrodynamics (SPH), while the
Eulerian method is the astrophysical hydrocode ZEUS.  In future work
we use this numerical calibration in the interpretation of
self-gravitating MHD models computed with ZEUS. 

\subsection{SPH}
\label{subsec:method-SPH}
SPH is a Lagrangian, particle-based scheme to solve the equations of
hydrodynamics.  The fluid is represented  by an ensemble of
particles, each carrying mass, momentum, and thermodynamical
properties.  Fluid properties at any point are obtained by averaging
over a set of neighboring particles.  The time evolution of the fluid
is represented by the time evolution of the particles, governed by the
equation of motion and equations to implement hydrodynamic properties.
The technique can therefore be seen as an extension of the pure
gravitational $N$-body system. Excellent overviews of the method, its
numerical implementation, and some of its applications are given by
the reviews by Benz (1990) and Monaghan (1992).  The code used here
derives from a version originally developed by Benz (1990).  It
includes a standard von~Neumann-type artificial viscosity (Monaghan \&
Gingold 1983) with the parameters $\alpha_v = 1$ and $\beta_v = 2$ for
the linear and quadratic terms.  The system is integrated in time
using a second-order Runge-Kutta-Fehlberg scheme, allowing individual
time steps for each particle.  Furthermore, the smoothing volume over
which hydrodynamic quantities are averaged in the code is freely
adjustable in space and time such that the number of neighbors for
each particle remains approximately fifty. 

SPH can resolve very high density contrasts because it increases the
particle concentration, and thus the effective spatial resolution, in
regions of high density, making it well suited for computing collapse
problems.  Conversely, it resolves low-density regions poorly. Shock
structures  tend to be broadened by the averaging kernel in the
absence of adaptive techniques.  It is also very difficult to include
magnetic fields in the algorithm.  SPH can be run on the
special-purpose hardware device GRAPE (Sugimoto et al.~1990, Ebisuzaki
et al.~1993; and also Umemura et al.~1993, Steinmetz 1996), which
allows supercomputer-level calculations to be done on a normal
workstation. As we concentrate on subregions inside molecular clouds
of much larger extent, we use periodic boundary conditions, as
implemented by Klessen (1997) on GRAPE.

The correct numerical treatment of gravitational collapse requires the
resolution of the local Jeans mass at every stage of the collapse
(Bate \& Burkert 1997).  In the current code, once an object with
density beyond the resolution limit of the code has formed in the
center of a collapsing gas clump it is replaced by a `sink' particle
(Bate, Bonnell, \& Price 1995).  This particle has a fixed radius on
the order of the Jeans length at the threshold density. We set this
density to be $10^4$ times the average density in the simulation,
which roughly corresonds to the maximum resolvable density contrast.  The
sink particle inherits the combined mass of the replaced SPH
particles, as well as their linear and angular momenta.  It has the
ability to accrete further SPH particles from its infalling gaseous
envelope, which are then removed from the computation. Adequately
replacing high-density cores and keeping track of their further
evolution in a consistent way prevents the time step from becoming
prohibitively small. We are thus able to follow the collapse of a
large number of cores until the overall gas reservoir becomes
exhausted.

\subsection{ZEUS-3D}
\label{subsec:method-ZEUS}
ZEUS-3D is a well-tested, Eulerian, finite-difference code (Stone \&
Norman 1992, Clarke 1994).  It uses second-order van Leer (1977)
advection, and resolves shocks using von Neumann artificial viscosity.
Self-gravity is implemented via an FFT-solver for Cartesian
coordinates.  It also includes magnetic fields in the
magnetohydrodynamic approximation.  For the models discussed here, we
use a three-dimensional, periodic, uniform, Cartesian grid.  This
gives us equal resolution in all regions, and allows us to resolve
shocks well everywhere.  On the other hand, collapsing regions cannot
be followed to scales less than one or two cells.
 
We must again consider the resolution required for gravitational
collapse.  For a grid-based simulation, the criterion given by
Truelove et al.~(1997) holds. Equivalent to the SPH resolution
criterion, the mass contained in one grid zone has to be smaller than
the local Jeans mass throughout the computation. Applying this
criterion strictly would limit our simulations to the very first
stages of collapse, as we have not implemented anything like sink
particles in ZEUS. We have therefore extended our models beyond the
point of full resolution of the collapse, as we are primarily
interested in the formation of collapsed regions, but not their
subsequent evolution.  Thus, in the ZEUS models, the fixed spatial
resolution of the grid implies that strongly collapsed cores have a
larger cross-section than appropriate for their mass. In encounters
with shock fronts the probability for these cores to get destroyed or
lose material is overestimated.  Cores simulated with ZEUS are
therefore more easily disrupted than they would be physically. SPH, on
the other hand, underestimates the disruption probability, because
sink particles cannot lose mass or dissolve again once they have
formed. The physical result is thus {\em bracketed} by these two
numerical methods (also see \S~\ref{sec:collapse}).

\subsection{Models}
\label{subsec:models}
We perform our computations using normalized units. The considered
volumes are cubes with side $L = 2$, extending from -1 to 1, which are
subject to periodic boundary conditions in every direction. The total
mass in the box is $M=1$, therefore the uniform initial density is
$\rho_0 = 1/8$. We use an isothermal equation of state, with sound
speed $c_{\rm s} = 0.1$, chosen to set the number of thermal Jeans
masses contained in the box to $N_{\rm J} = 64$.  Time is measured in
units of the initial global free-fall time of the system. 

To generate and maintain turbulent flows we introduce Gaussian
velocity fluctuations with power only in a narrow interval $k-1 \leq
|\vec{k}| \leq k$, where $k = L/\lambda_d$ counts the number of
driving wave lengths $\lambda_d$ in the box.  This offers a simple
approximation to driving by mechanisms that act on that scale.
Comparing runs with different $k$ will then give some information on
how, for example, turbulence driven by large-scale shearing motions
might differ from turbulence driven by low-mass protostars.  We set up
the initial velocity field as described in Mac Low et al.~(1998), with
perturbations drawn from a Gaussian random field determined by its
power distribution in Fourier space, following the usual procedure.
For each three-dimensional wave number $\vec{k}$ we randomly select an
amplitude from a Gaussian distribution around unity and a phase
between zero and $2\pi$.  We then transform the resulting field back
into real space to get a velocity component in each zone, and multiply
by the amplitude required to get the desired initial root mean square
(rms) velocity on the grid.  We repeat this for each velocity
component independently to get the full velocity field.

To drive the turbulence, we then normalize this fixed pattern to
produce a set of perturbations $\delta\vec{\nu}(x,y,z)$, and at every
time step add a velocity field $\delta\vec{v}(x,y,z) = A
\delta\vec{\nu}$ to the velocity $\vec{v}$.  The amplitude $A$ is
chosen to maintain constant kinetic energy input rate $\dot{E}_{\rm
in} = \Delta E / \Delta t$.  For a compressible flow with a
time-dependent density distribution, we maintain a constant energy
input rate by solving a quadratic equation in the amplitude $A$ at
each time step, as discussed in Mac~Low (1999). In dynamical
equilibrium, the driving luminosity $\dot{E}_{\rm in}$ equals the rate
of turbulent energy dissipation.  To estimate the input rate necessary
to reach and maintain a certain equilibrium level of the kinetic
energy we use equation~7 of Mac~Low (1999).  We find that this
equation underestimates the driving energy needed to maintain the SPH
models at a specific equilibrium kinetic energy by 20--30\% for
reasons that we do not yet fully understand.  Comparisons with other
techniques will probably be required to resolve this discrepancy.  We
drive the SPH models somewhat harder to compensate, as can be seen in
table~\ref{tab:models} by comparing $\dot{E}_{\rm in}$ for SPH and
ZEUS models with equivalent driving wave length $k_{\rm drv}$ and
$E^{eq}_{\rm kin}$.  Dynamical equilibrium is reached typically
within one global shock crossing time $t=L/\langle v \rangle$. The
equilibrium value is determined to an accuracy of better than 10\%.
Keeping the energy input unaltered we then switch on self-gravity with
gravitational constant $G=1$, and allow the evolution to proceed. This
defines $t=0$ in our models. Their most important properties are
summarized in table~\ref{tab:models}.

\subsection{Scaling}
\label{subsec:scaling}
The dynamical behavior of isothermal self-gravitating gas is scale
free and depends only on the ratio between potential energy and
kinetic energy (including thermal energy).  We can scale our models to
physical units with a mass scale of the thermal Jeans mass $M_{\rm J}$
given by equation~(\ref{eqn:jeans-mass}), a length scale given by the
Jeans length $\lambda_{\rm J}$ derived from
equation~(\ref{eqn:jeans-wave-number}), and a time scale given by the
free-fall time scaled from equation~(\ref{eqn:free-fall-time})
\begin{equation}
\tau_{\rm ff} = (0.34 \mbox{ Myr}) \left(\frac{n}{10^4 
                                 \mbox{ cm}^{-3}}\right)^{-1/2}, 
\end{equation}
where $G = 6.67 \times 10^{-8}$ cm$^3$ g$^{-1}$ s$^{-2}$,  and the number
density is taken as $n = \rho/\mu$, with $\mu = 2.36\, m_{\rm H}$.  The 120
thermal Jeans masses in our simulation cubes then correspond to
\begin{equation}
M = (413 M_{\odot}) 
    \left(\frac{c_{\rm s}}{0.2 \mbox{ km s}^{-1}}\right)^3
    \left(\frac{n}{10^4 \mbox{ cm}^{-3}}\right)^{-1/2},
\end{equation}
where a sound speed $c_{\rm s} = 0.2$ km s$^{-1}$ corresponds to a
temperature $T = 11.4$ K with the value of $\mu$ we use.
Finally we may compute the size of our cube by noting that the Jeans
length in our computational units is $\lambda_{\rm J} = 0.1 \sqrt{8
\pi} \approx 0.501$, while the size of the cube is $L = 2$, so that in
physical units
\begin{equation}
L = (0.89  \mbox{ pc})
    \left(\frac{c_{\rm s}}{0.2 \mbox{ km s}^{-1}}\right)
    \left(\frac{n}{10^4 \mbox{ cm}^{-3}}\right)^{-1/2}.
\end{equation}
  
As an example, let us consider a dark cloud like Taurus with $n({\rm
H}_2) \approx 10^2\,{\rm cm}^{-3}$, and $c_{\rm s} \approx 0.2$
km~s$^{-1}$.  Then our simulation cube holds a mass $M = 4.1 \times
10^3\,$M$_{\odot}$ and has a size $L = 8.9\,$pc. The time unit
(free-fall time scale) is $\tau_{\rm ff}=3.4\,$Myr, and the average
{\em thermal} Jeans mass for the homogeneous distribution follows as
$M_{\rm J} = 65\,$M$_{\odot}$.  Another example would be a dense cloud
forming massive stars such as the BN region in Orion, with $n({\rm
H}_2) \approx 10^5\,$cm$^{-3}$ and $c_{\rm s} \approx 0.2$
km$\,$s$^{-1}$.  Here the simulated cube holds a mass of $M =
130\,$M$_{\odot}$ and is of size $L = 0.28\,$pc. The time unit is now
$\tau_{\rm ff}= 0.1\,$Myr, and the thermal Jeans mass is $M_{\rm J} =
2.1\,{\rm M}_{\odot}$. (Note again, that in the spherical definition
the Jeans mass is smaller by a factor $\pi/6$.) 

\section{Local vs. Global Collapse}
\label{sec:collapse}

In this section we begin by showing numerical results that suggest
that local collapse can occur in turbulent self-gravitating media even
if the kinetic energy contained in the system is sufficient to
stabilize it on global scales (\S~\ref{subsec:local-collapse-1}).  The
strong shocks ubiquitous in supersonic turbulence compress small
regions sufficiently that the turbulence can no longer support them.
We then consider what promotes or prevents this process
(\S~\ref{subsec:local-collapse-2}), and the importance of turbulent
collapse in real molecular clouds (\S~\ref{subsec:real-clouds}).

\subsection{Local Collapse in a Globally Stable Region}
\label{subsec:local-collapse-1}

We compute models with both SPH and ZEUS in which the turbulence is
driven at strengths above and below the critical value needed to
prevent gravitational collapse according to the analytic predictions
of \S~\ref{sec:jeans}.  The models can be characterized by two
parameters, the kinetic energy before gravity is turned on, and the
typical driving wave number $k$ at which energy is injected (see
\S~\ref{subsec:models}).  We define an {\em effective turbulent Jeans
mass} $\langle M_{\rm J}\rangle_{\rm turb}$ by substituting $c_{\rm
s}^2 \longrightarrow c_{\rm s}^2 + 1/3 \, \langle v^2 \rangle$ for the
thermal sound speed $c_{\rm s}$ in equation~(\ref{eqn:jeans-mass})
where we approximate the rms velocity of the flow $\langle v^2
\rangle$ by $2E_{\rm kin}/ M$.  We do simulations with $\langle M_{\rm
J}\rangle_{\rm turb}$ of 0.6, 3.2, and 18.2. These values have to be
compared to the total system mass $M\equiv 1$ in order to determine
whether global stability is reached. Note that our definition of the
Jeans mass uses the mean density in the simulations. This is
equivalent to examining the collapse properties of isolated gas
cubes. In infinite media (local) density contrasts should be used
instead. For this reason, the quoted turbulent Jeans masses are lower
limits to the true critical values for support against gravitational
collapse. The true stabilizing effect of turbulence on large scales is
{\em stronger} than indicated from merely comparing $\langle M_{\rm
J}\rangle_{\rm turb}$ with the total mass in the system.

\begin{figure}[h]
\unitlength1.0cm
\begin{picture}(16,22)
\put(-0.5,0){\epsfbox{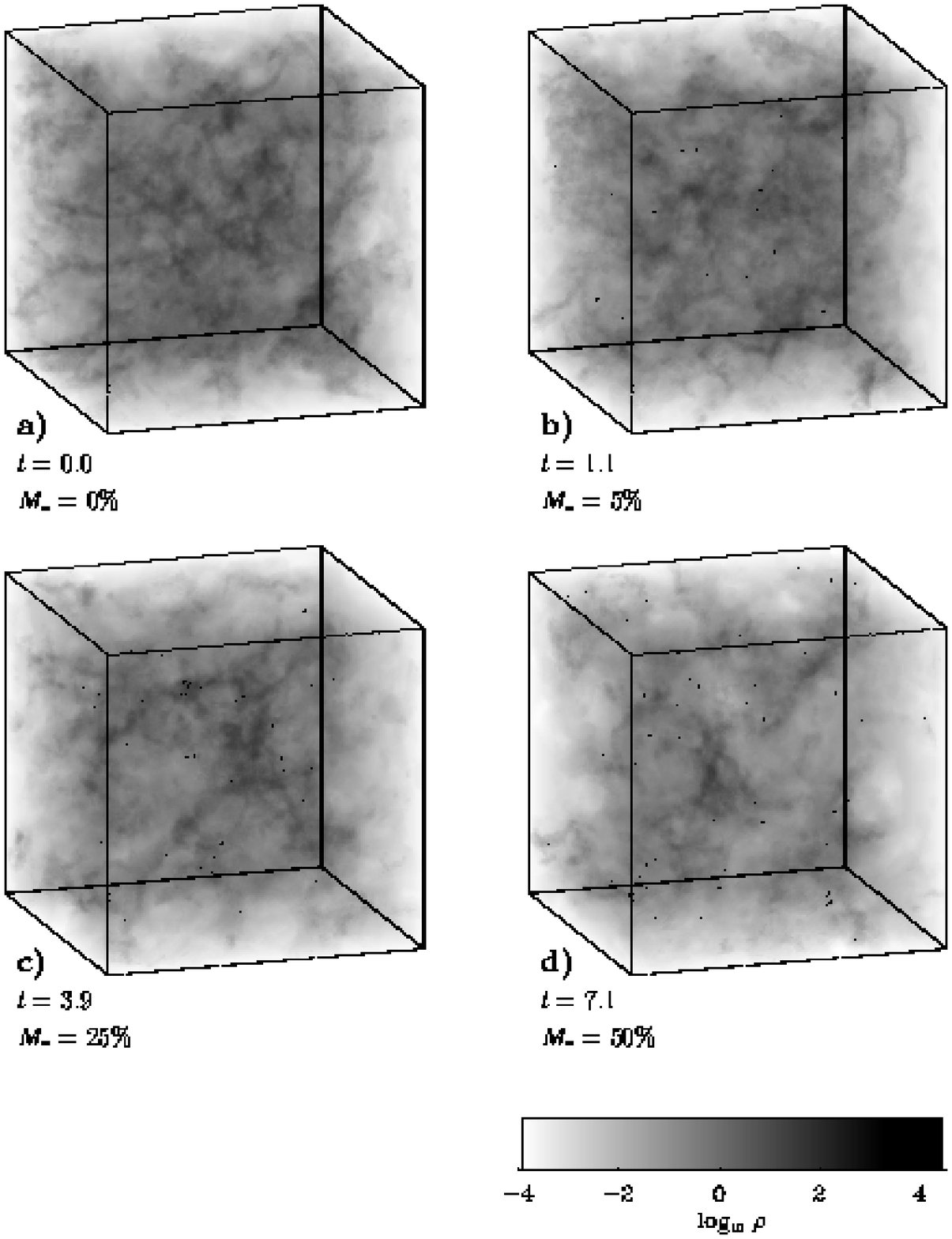}}
\end{picture}
\caption{\label{fig:3D-cubes}
SPH density cubes for model ${\cal B}2h$, which is driven in the
interval $3\le k \le 4$, shown (a) at the time when
gravity is turned on, (b) when the first dense cores are formed and
have accreted $M_* = 5$\% of the mass, (c) when the mass in dense cores is
$M_* = 25$\%, and (d) when  $M_* = 50$\%. Time is measured
in units of the global system free-fall time scale   $\tau_{\rm
ff}$. }
\end{figure}

We find that {\em local} collapse occurs even when the turbulent
velocity field carries enough energy to counterbalance gravitational
contraction on global scales.  This confirms the results of
two-dimensional (2D) and low-resolution ($64^3$) 3D computations with
and without magnetic fields by V\'azquez-Semadeni et al.\ (1996).  An
example of local collapse in a globally supported cloud is given in
figure~\ref{fig:3D-cubes}.  It shows a sequence of 3D density cubes of
the SPH model ${\cal B}2h$ which is driven in the wave length interval
$3\le k \le 4$ so that the turbulent Jeans mass $\langle M_J
\rangle_{\rm turb} = 3.2$.  The first cube shows the system at
$t=0.0$. Hydrodynamic turbulence is fully established but gravity has
not yet been included in the computation.  (Note again that time is
measured in units of the global free-fall time of the system
$\tau_{\rm ff}$ and the zero-point is set when gravity is switched
on).  The second cube shows the system at a time $t=1.1$. Density
fluctuations generated by supersonic turbulence in converging and
interacting shock fronts that locally exceed the Jeans limit begin to
contract.  The central regions of these high-density clumps have
undergone sufficient gravitational contraction to be identified as
collapsed cores. Numerically, in the SPH code they have been replaced
with sink particles. There are altogether twelve cores containing $M_*
= 5$\% of the total gas mass in the system. At $t=3.9$ the number of
dense embedded cores has grown to 46 and they account for 25\% of the
mass. At $t=7.1$ roughly 50\% of the gas mass is accreted onto 53
dense cores. The first cores form in small groups randomly dispersed
throughout the volume. Their velocities directly reflect the turbulent
velocity field of the gas they are created from, in which they are
still embedded, and from which they continue to accrete. However, as
more and more mass accumulates on the cores the gravitational
interaction between the cores themselves increasingly determines their
dynamical evolution. The core cluster begins to behave more like a
collisional $N$-body system, in which close encounters are dynamically
important.  

Local collapse in a globally stabilized cloud is not predicted by the
analytic models described in Sec.~\ref{sec:jeans}. For the parameters
of the models presented here, the dispersion relation
equation~\ref{eqn:bonazzola-dispersion-rel} forbids gravitational
contraction at any scale.  However, this equation was derived under
the assumption of incompressibility.  The presence of shocks in
supersonic turbulence drastically alters the result, as was first
noted by Elmegreen (1993) and studied numerically by
V\'azquez-Semadeni et al.\ (1996).  The density contrast in isothermal
shocks scales quadratically with the Mach number, so the shocks driven
by supersonic turbulence create density enhancements with $\delta \rho
\propto {\cal M}^2$, where ${\cal M}$ is the {\em rms} Mach number of
the flow. In such fluctuations the local Jeans mass is {\em
decreased} by a factor of $\cal M$ and therefore the likelihood for
gravitational collapse {\em increased}.

To test this explanation numerically, we designed a test case driven
at short enough wave length and high enough power to support even
fluctuations with $\delta \rho \propto {\cal M}^2$, and ran it with
both codes.  To ensure sufficient numerical resolution for these
models, ${\cal B}5$ and ${\cal D}5$, we computed a subvolume of mass
$M = 0.25$ with reduced sound speed $c_{\rm s} = 0.05$ driven at
wave number $k=9-10$. This is equivalent to an {\em effective} driving
wave number $k=39-40$ on the regular cube ($M=1$, $c_{\rm s} = 0.1$).
Within 20 $\tau_{\rm ff}$ neither of these models show signs of
collapse.  All the other globally supported models with less extreme
parameters that we computed did form dense cores during the course of
their evolution, supporting our hypothesis that local collapse is
caused by the density fluctuations resulting from supersonic
turbulence.

The two numerical methods that we use are complementary, as discussed
in \S~\ref{sec:method}.  SPH is a particle based, Lagrangian scheme.
It resolves regions of high density well, and the use of sink
particles makes it straightforward to define dense cores, but it does
not resolve shocks well.  Once a collapsing region passes beyond the
density threshold and is converted into a sink particle, it cannot be
destroyed.  It continues to accrete matter from its surroundings and
to interact gravitationally with other cores. This {\em overestimates}
the survival probability of collapsing, Jeans-unstable fluctuations.
ZEUS, on the other hand, is an Eulerian grid method, well suited for
resolving shocks, but worse at modeling gravitational collapse.  Due
to the fixed grid, it overestimates the volume of collapsed cores,
leading to an enhanced cross-section to destructive processes such as
tidal interactions between cores, or the perturbations of passing
shock fronts. Hence, the probability for core formation and survival
in the turbulent environment is {\em underestimated}.  The real
behavior of self-gravitating, turbulent gas lies in between, {\em
bracketed} by the two methods which we apply here.  Both methods show
local collapse occurring in globally stabilized clouds.

\begin{figure}[h]
\unitlength1.0cm
\begin{picture}(16,20)
\put(-0.5,0){\epsfbox{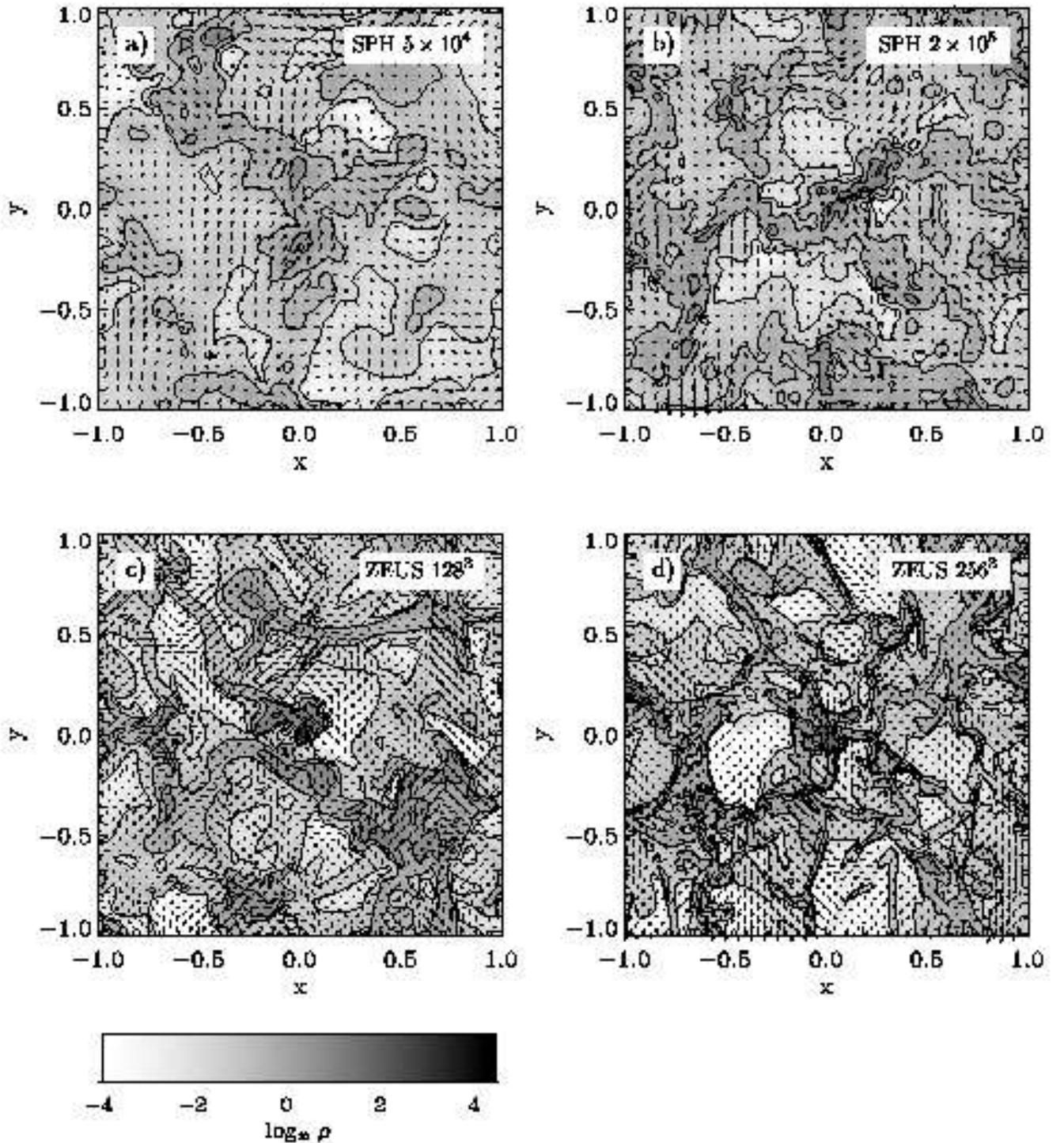}}
\end{picture}
\caption{\label{fig:2D-cuts} Comparison of 2D density slices through
3D models with identical physical parameters ($\langle M_{\rm
J}\rangle_{\rm turb} = 3.2$ and $k=3-4$) computed with different
numerical methods and resolution: SPH models (a) ${\cal B}2$ and (b)
${\cal B}2h$ with $50\,000$ and $200\,000$ particles, and ZEUS models
(c) ${\cal D}2$ and (d) ${\cal D}2h$ with $128^3$ and $256^3$ grid
cells. For further details see table~\ref{tab:models}. To allow for
comparison, the time is chosen such that the mass accreted onto dense
cores is $M_* = 5$\%. Density is scaled logarithmically with the
separation of contour levels being one decade. Each cut is centered on
the density maximum in the simulation.  In SPH, the density
distribution has been interpolated onto a uniform grid using kernel
smoothing.  The arrows indicate the velocity components {\em in} the
plane of section.}
\end{figure}

Figure~\ref{fig:2D-cuts} illustrates this point by comparing 2D slices
through 3D models which are run by the two different codes with
similar turbulent driving power spectra at both medium and high
resolution.  Each slice is centered on the densest core on the grid.
(Because we use periodic boundary conditions, we are free to shift the
window across the simulated volume in any direction.  These boundaries
do not introduce artificial perturbations.)  Figures
\ref{fig:2D-cuts}a and b show the SPH models ${\cal B}2$ and ${\cal
B}2h$ with $50\,000$ and $200\,000$ particles, while figures
\ref{fig:2D-cuts}c and d show the ZEUS models ${\cal D}2$ and ${\cal
D}2h$ with $128^3$ and $256^3$ grid zones.  We use a new realization
of the initial conditions with the same statistical properties for
each of these models, so there is no expectation that they will have
identical structures, only that they will have similar typical
structures.  The roundish appearance of structures in the SPH models,
especially at lower resolution, stems from the smoothing intrinsic to
the SPH algorithm.  The Lagrangian nature of the scheme leads to high
spatial resolution in high-density regions but degraded resolution in
low-density regions where particles are sparse. Conversely, ZEUS does
very well at modeling the shock and void structure, especially in the
high-resolution model ${\cal D}2h$, but the dense collapsed cores are
underresolved.  The shocks and filaments clearly resolved by the ZEUS
model are also present in the SPH model, but tend to be rather smeared
out by the lack of resolution in the lower density regions.
Nevertheless, {\em all} the images clearly indicate the presence of
strong shocks which sweep up gas into gravitationally collapsing
regions.

\subsection{Promotion and Prevention of Local Collapse}
\label{subsec:local-collapse-2}

The total mass and lifetime of a fluctuation determine whether it will
actually collapse.  Roughly speaking, the lifetime of a clump is
determined by the interval between two successive passing shocks: the
first creates it, while if the second is strong enough, it disrupts
the clump again if it has not already collapsed (Klein, McKee \&
Colella 1994, Mac Low et al.\ 1994).  If its lifetime is long enough,
a Jeans unstable clump can contract to sufficiently high densities to
effectively decouple from the ambient gas flow. It then becomes able
to survive the encounter with further shock fronts (e.g. Krebs \&
Hillebrandt 1983), and continues to accrete from the surrounding gas,
forming a dense core.  The weaker the passing shocks, and the greater
the separation between them, the more likely that collapse will occur.
Equivalently, weak driving and long typical driving wave lengths
enhance collapse.  The influence of the driving wave length is enhanced
because individual shocks sweep up more mass when the typical
wave length is longer, so density enhancements resulting from the
interaction of shocked layers will have larger masses, and so are more
likely to exceed their local Jeans limit.  Turbulent driving
mechanisms that act on large scales will produce large coherent
structures (filaments of compressed gas with embedded dense cores) on
relatively short time scales compared to small-scale driving even if
the total kinetic energy in the system is the same.

\begin{figure}[h]
\unitlength1.0cm
\begin{picture}(16,10)
\put(-0.5,0){\epsfbox{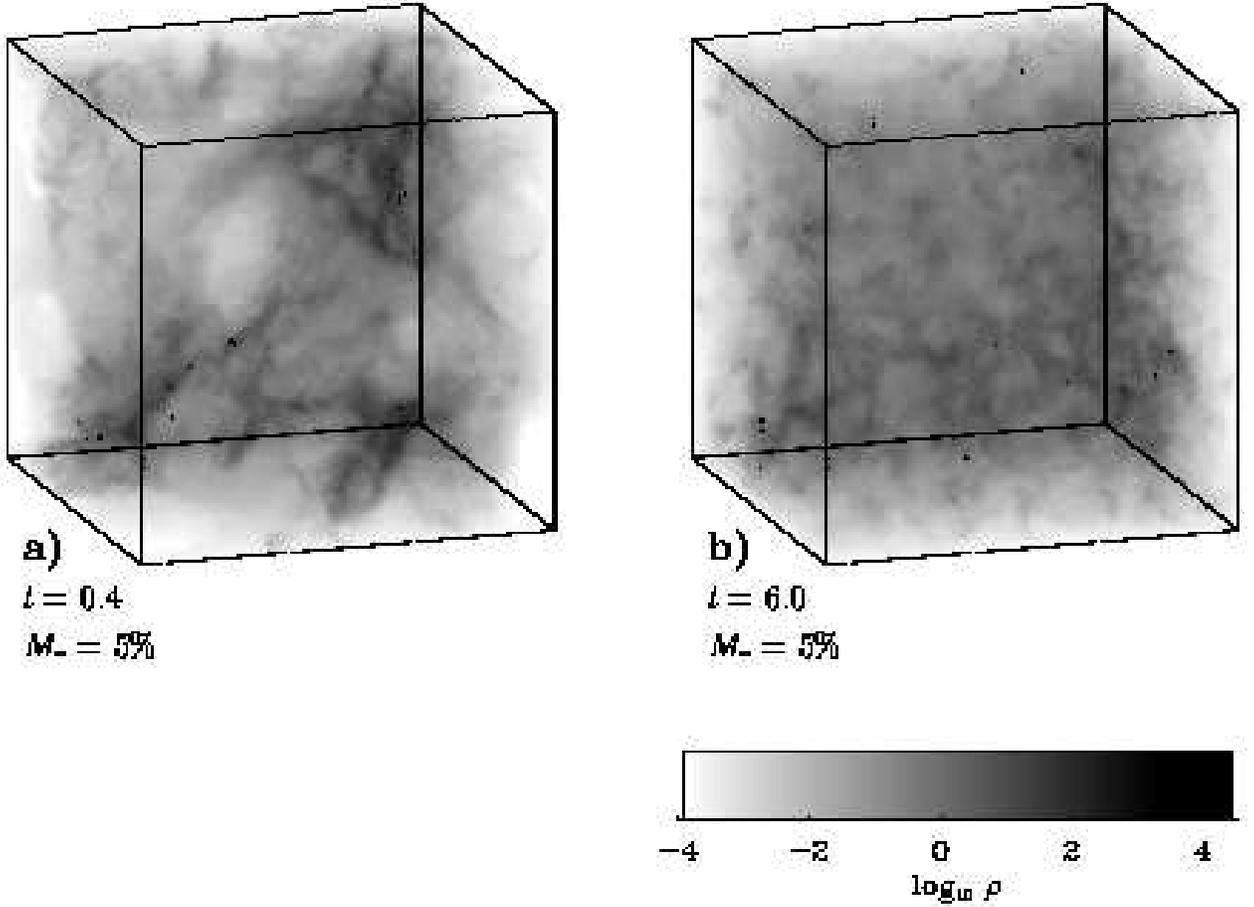}}
\end{picture}
\caption{\label{fig:3D-cubes-1..2+7..8} Density cubes for models (a)
${\cal B}1h$ ($k=1-2$) and (b) ${\cal B}3$ ($k=7-8$) at dynamical
stages where the core mass fraction is $M_* = 5$\%. Compare these
figures with figure~\ref{fig:3D-cubes}b. Together they show the
influence of different driving wave lengths for otherwise identical
physical parameters. Note the different visual appearance of the
systems and the different times at which $M_* = 5\%$ is reached. }
\end{figure}

We demonstrate the effect of the driving wave length in
figure~\ref{fig:3D-cubes-1..2+7..8}, which compares SPH model ${\cal
B}1h$ with driving wave numbers $k=1-2$ to model ${\cal B}3$ driven
with $k=7-8$ at a time when sink particles have accreted 5\% of the
gas mass.  (These density cubes can be directly compared with
figure~\ref{fig:3D-cubes}b, which shows the intermediate case $k=3-4$
at the same evolutionary stage.)  Note the difference in the
morphology of the density structures.
Figure~\ref{fig:3D-cubes-1..2+7..8}a is dominated by {\em one} large
shock front that traverses the volume, which is the sole site of core
formation. On the other hand, the density structure in model ${\cal
B}3$ (figure \ref{fig:3D-cubes-1..2+7..8}b) is far more
homogeneous, without any large-scale structure.  Cores form alone,
randomly dispersed throughout the volume.  This comparison is
discussed below in \S~\ref{subsec:clustered-isolated-SF}.

The influence of driving strength and wave length on local collapse
can be examined by measuring the amount of mass accreted onto
collapsed regions over time in each model.  In the SPH models, this
can be computed quite simply by adding up the masses of the sink
particles at each time.  As ZEUS does not include sink particles, we
instead employ a modified version of the {\sc Clumpfind} method
(Williams, de Geus, \& Blitz 1994; see also appendix 1 in Klessen \&
Burkert 2000).  In this routine, clumps are defined as regions of
connected zones whose densities lie above a certain threshold.  In
order to be able to use {\sc Clumpfind} on models as large as $256^3$
zones, we replaced the inefficient clump identification routines with
an algorithm based on the dilation operators implemented in IDL.  We
use two criteria to separate collapsed cores from shock-generated
fluctuations.  First, we require the average density of those cores to
exceed the mean value expected for 
isothermal shocks, $\rho > {\cal M}^2\rho_0$. Here, ${\cal M}$ is the
rms Mach number and $\rho_0$ is the mean density.  Second, we count
only  fluctuations for which the potential energy exceeds the
kinetic energy, $E_{\rm kin}^{core} < |E_{\rm pot}^{core}|$, and which
are more massive than the local Jeans mass, $M_* > M_{\rm
J}^{core}(\rho)$.  We use logarithmic density contours instead of
linear ones in order to get a wide enough density range so that most
detected clumps consist of more than one cell.  However, we also
accept single high-density cells as cores. These are common at late
stages of the evolution, when the envelopes of cores have been removed
by further shock interactions and only the collapsed centers
remain. In this case the we use the ratio of  potential to internal 
energy as the criterion for collapse. 

A figure of merit that we can use to examine the effect of driving
strength and wave length is the time $t_{5\%}$ needed to sweep up 5\%
of the mass into compact cores.  Table~\ref{tab:models} describes
several sequences of models identical except for their driving wave
length: the high and medium resolution SPH models ${\cal A}1-{\cal
A}3$, ${\cal B}1-{\cal B}4$ and ${\cal C}2$, and the low, medium, and
high resolution versions of the ZEUS models ${\cal D}1-{\cal D}3$.
Comparison of the values of $t_{5\%}$ for these models shows that
collapse and accretion occurs more rapidly for models with larger
driving wave length (smaller driving wave numbers and larger typical
scales).  Comparison of the SPH models with $k = 3-4$ shows that
stronger driving also delays collapse.

Models run with the same driving strength at different resolutions and with
the two different codes can be compared to determine the level of numerical
convergence, and the effect of the different algorithms.  Comparison of SPH
models ${\cal B}2\ell$ to ${\cal B}2h$ shows that a change of linear
resolution of 2.2 yields a change in $t_{5\%}$ of only 12.5\%.  Similarly
comparison of the $128^3$ zone to the $256^3$ zone resolution ZEUS models
shows better than 10\% agreement, except for high wave number driving,
where the disagreement is still less than 25\%.  Comparison of the ZEUS and
SPH models with $k = 3-4$ driving of the same strength also shows better than
25\% quantitative agreement.  The 2D cuts through medium and high resolution
models with both codes with the same driving wave length ($k=3-4$) and driving
strength ($\langle M_{\rm J}\rangle_{\rm turb} = 3.2$) presented in
figure~\ref{fig:2D-cuts} visually demonstrate the level of morphological
agreement.  {\em We emphasize that the qualitative result that local collapse 
occurs at a rate dependent on the driving wave length and strength is
recovered at all resolutions and with both codes.}

\begin{figure}[h]
\unitlength1.0cm
\begin{picture}(16,8)
\put(-1.00, -11.00){\epsfbox{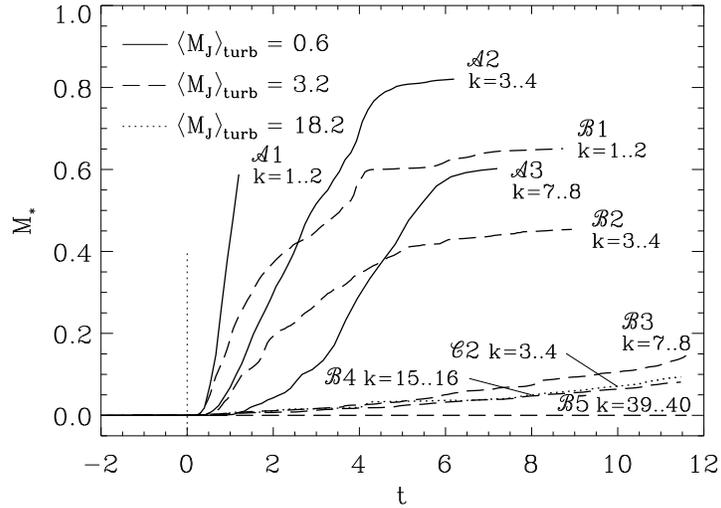}}
\end{picture}
\caption{\label{fig:accretion-history} Fraction of mass $M_*$ in dense
cores as function of time.  All models are computed using SPH with
sink particles replacing dense, collapsed cores. The different models
are indicated in the figure, details can be found in
table~\ref{tab:models}. The figure shows how the efficiency of local
collapse depends on the scale and strength of turbulent driving. }
\end{figure}

A more detailed understanding of how local collapse proceeds comes
from examining the full time history of accretion for each model.
Figure \ref{fig:accretion-history} shows the accretion history for
three sets of SPH models.  For each set of models, the driving
strength is held constant while the effective driving wave length is
varied, showing the pronounced effect of the wave length at equal
driving strength.  At the extreme, if the driving is at wave lengths
below the Jeans wave length of the shocked layers local collapse does
not occur (model ${\cal B}5$).  The ${\cal A}$ models have lower
driving strength than the ${\cal B}$ and ${\cal C}$ models,
demonstrating the effect of driving strength at each driving wave
length.  

The cessation of strong accretion onto cores occurs long before all
gas has been accreted.  This appears to be because the time that dense
cores spend in shock-compressed, high-density regions decreases with
increasing driving wave number and increasing driving strength.  In
the case of long wave length driving, cores form coherently in
high-density regions associated with one or two large shock fronts
that can accumulate a considerable fraction of the total mass of the
system. The overall accretion rate is high and cores spend suffient
time in this environment to accrete a large fraction of the total mass
in the region.  Any further mass growth has to occur from chance
encounters with other dense regions.  In the case of short wave length
driving, the network of shocks is tightly knit. Cores form in shock
generated clumps of small masses because individual shocks are not
able to sweep up much matter. Furthermore, in this rapidely changing
environment the time interval between the formation of clumps and
their destruction is short.  The period during which individual cores
are located in high-density regions where they are able to accrete at
high rate is short as well. Altogether, the global accretion rates are
small and begin to saturate at lower values of $M_*$ as the driving
wave length is decreased. 

\begin{figure}[h]
\unitlength1.0cm
\begin{picture}(16,9)
\put(-3.40,-12.40){\epsfbox{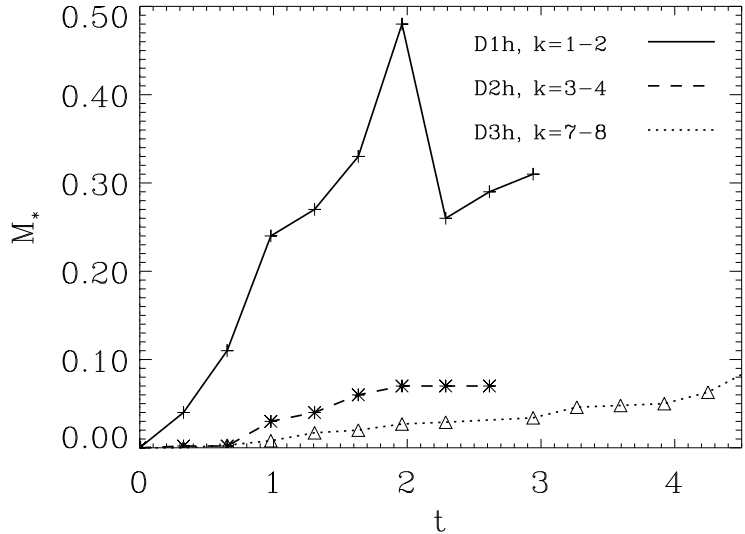}}
\end{picture}
\caption{\label{fig:accretion-history-ZEUS} Mass fraction $M_*$ in
dense cores as function of time for the three $256^3$ ZEUS models
driven with $k=1-2$ (solid, crosses), $k=3-4$ (dashed, stars) and
$k=7-8$ (dotted, triangles). $M_*$ is the sum of all cores found by
{\sc Clumpfind} as discussed in the text. Note that the method
identifies cores only after gravity is turned on, i.e. for $t >
0.0$. }
\end{figure}

Figure \ref{fig:accretion-history-ZEUS} shows the accretion history
for the three $256^3$ ZEUS models, ${\cal D}1h$ to ${\cal D}3h$.  The
fractional core mass $M_*$ in the model with large scale driving
(${\cal D}1h$) is strongly affected by the large shocks that run
through the volume.  At $t \approx 2.1$, for example, a shock destroys
the most massive core, so $M_*$ drops suddenly. Between successive
shock passages, the cores recover, so they accrete a substantial mass
fraction over the run.  Models ${\cal D}2h$ and ${\cal D}3h$ with
$k=3-4$ and $k=7-8$ display a steady mass growth similar to the SPH
models. The more frequent shocks in these models reduce the
accretion rate by stripping away material from the vicinity
of the central high-density zones. These isolated zones do not lose
mass from shock encounters, but are subject to numerical clipping, so
the measured fraction is, as explained before, a lower limit to the
actual accretion fraction.  A clear indication of local collapse is
once again seen.

\begin{figure}[h]
\unitlength1.0cm
\begin{picture}(16,7.5)
\put(-1.10,-11.00){\epsfbox{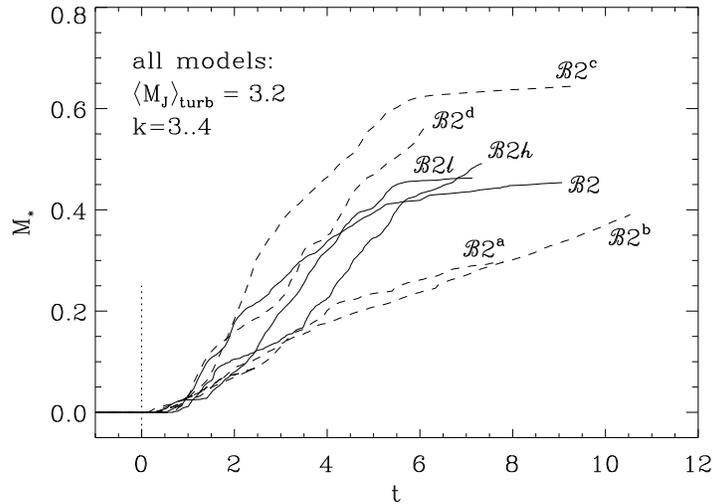}}
\end{picture}
\caption{\label{fig:accretion-history-resolution-study-SPH}
Study of resolution and statistical variation of the core mass
fraction $M_*$ over time for SPH models with turbulent Jeans mass
$\langle M_{\rm J} \rangle_{\rm turb} = 3.2$ and $k=3-4$.  The
low-resolution model ${\cal B}2\ell$ has $20\,000$ particles, for
medium-resolution model ${\cal B}2$ this number is $50\,000$, and for
the high-resolution model ${\cal B}2h$ it is $200\,000$. Model ${\cal
B}2$ has been repeated four times with different realizations of the
driving field. The alternative models ${\cal B}2^{a}$ to ${\cal
B}2^{d}$ are indicated by dotted lines. Note the large variance
effect.}
\end{figure}

To further investigate the influence of numerical resolution,
figure~\ref{fig:accretion-history-resolution-study-SPH} compares the
time history of accretion for SPH models with varying particle
numbers, but identical turbulent Jeans mass $\langle M_{\rm
J}\rangle_{\rm turb} = 3.2$ and driving wave numbers $3\le k \le 4$
(see table~\ref{tab:models}). The difference in effective linear
resolution (cube root of the particle number) between ${\cal B}2\ell$
and ${\cal B}2h$ is 2.2.  We also had to distinguish the effects of
statistical variance from the effects of resolution. To do this, we
repeated the intermediate resolution simulation ${\cal B}2$ four more
times, varying only the random seeds used to generate the Gaussian
fields (models ${\cal B}2^{a}$--${\cal B}2^{d}$, dashed lines).  We
actually find stronger variation between the different models at the
same resolutions than between models at different resolutions,
suggesting that numerical diffusivity doesn't have as large an effect
as the natural statistical variation.  This is not surprising given
the stochastic nature of turbulent flows.  Protostellar cores form in
molecular clouds through a sequence of highly probabilistic events.
Especially at late times, their mass accretion is strongly influenced
by chaotic $N$-body dynamics (Klessen et al.~1998, Klessen \& Burkert
2000).  All models agree well at early times when initial local
collapse occurs, suggesting that we are well converged on our basic
result.  At late times, variations between the different models become
stronger.  These variance effects need to be kept in mind when
interpreting the accretion rates of individual models.  For the
ensemble average at late times we do not expect significant variations
at the different numerical resolutions which we study, though our
current set of calculations is not large enough to quantify this
statement.

\begin{figure}[h]
\unitlength1.0cm
\begin{picture}(16,8.5)
\put(-3.40,-12.40){\epsfbox{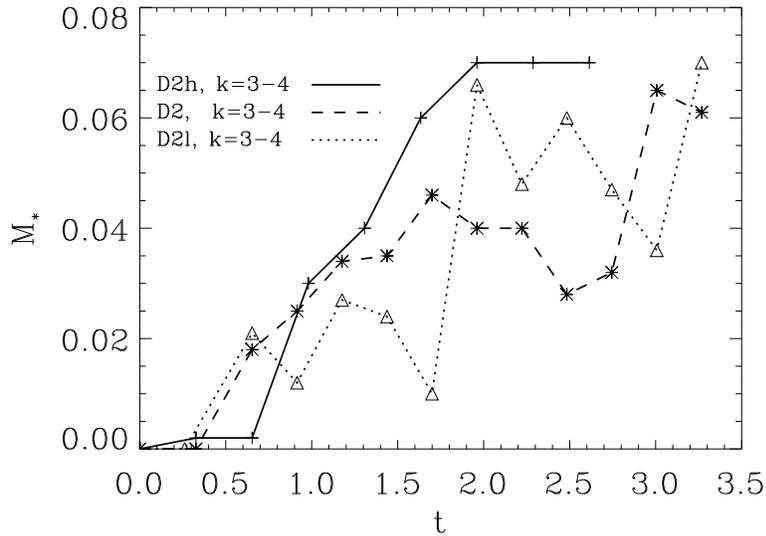}}
\end{picture}
\caption{\label{fig:accretion-history-resolution-study-ZEUS} 
Resolution study of core mass fraction $M_*$ as function of time
for ZEUS models with turbulent Jeans mass $\langle M_{\rm J}
\rangle_{\rm turb} = 3.2$ and driving wave number $k=3-4$. The models
have resolutions of $64^3$ (dotted), $128^3$ (dashed), and $256^3$
cells (solid). $M_*$ is computed using {\sc Clumpfind} as discussed in
the text.}
\end{figure}

Figure~\ref{fig:accretion-history-resolution-study-ZEUS} shows the
time history of accretion for ZEUS models with the same parameters as
the SPH models shown in
figure~\ref{fig:accretion-history-resolution-study-SPH}, and numerical
resolution increasing from $64^3$ to $256^3$ zones.  Strong
fluctuations in the lower resolution curves are caused by core
disruptions due to shocks, which cannot occur in the SPH models as
sink particles are never destroyed.  The fluctuations decrease with
increasing resolution because the cores have smaller cross section in
the high resolution models, and are thus less liable to be destroyed
by shocks. The high resolution model ${\cal D}2h$ shows a well defined
accretion behavior and reaches a saturation level at a mass fraction
of about 8\%, where the local free-fall time of the cores is roughly
equal to the time interval between two shock passages.  All three
models reach this level at least intermittently, suggesting it defines
a reasonably firm upper limit for these ZEUS models, and thus a lower
limit to the amount of mass that can actually be accreted under these
physical conditions, with the SPH models giving an upper limit.

\subsection{Application to Molecular Clouds}
\label{subsec:real-clouds}

The {\em global} star formation efficiency in normal molecular clouds
is usually estimated be of the order of a few per cent. Their life
times are typically thought to be a few$ \times 10^7$ years which is
equivalent to a few tens of their free-fall time $\tau_{\rm ff}$
(Blitz \& Shu 1980, Blitz 1993, Williams et al.~1999). It would be
consistent with these estimations if the mass fraction of protostellar
cores in our simulations remained below $M_* = 5$\% for
$~10\,\tau_{\rm ff}$. Indeed, as indicated in column 8 of
table~\ref{tab:models} local collapse can be slowed down considerably
in the case of small-scale driving.  However, if the hypothesis of
rapid molecular cloud evolution is correct (Ballesteros-Paredes et
al. 1999, Elmegreen 2000), the constraints on the driving scale and strength are
 substantially changed.
Furthermore, it needs to be noted that the
{\em local} star formation efficiency in molecular clouds can reach
very high values. For example, the Trapezium star cluster in Orion is
likely to have formed with an efficiency of about 50\% (Hillenbrand \&
Hartmann 1998).  In \S~\ref{subsec:clustered-isolated-SF} we will
argue that this apparent difference between the `clustered' and
`isolated' model of star formation can be explained in terms of the
properties of the underlying turbulent velocity field of the parental
gas. 

The energy dissipation scale in molecular clouds should also be
considered.  It was first shown by Zweibel \& Josafatsson (1983) that
ambipolar diffusion would be the most important dissipation mechanism
in typical molecular clouds with very low ionization fractions $x =
\rho_i/\rho_n$, where $\rho_i$ is the density of ions, $\rho_n$ is the
density of neutrals, and $\rho = \rho_i + \rho_n$.  
An ambipolar diffusion strength can be defined as
\begin{equation}
\lambda_{AD} = v_A^2 / \nu_{ni},
\end{equation}
where $v_A^2 = B^2/4\pi\rho_n$ approximates the effective Alfv\'en
speed for the coupled neutrals and ions if $\rho_n \gg \rho_i$, and
$\nu_{ni} = \gamma \rho_i$ is the rate at which each neutral is hit by
ions.  The coupling constant depends on the cross-section for
ion-neutral interaction, and for typical molecular cloud conditions
has a value of $\gamma \approx 9.2 \times
10^{13}$~cm$^3$~s$^{-1}$~g$^{-1}$ (e.g.\ Smith \& Mac Low 1997).
Zweibel \& Brandenburg (1997)  define an ambipolar diffusion
Reynolds number as
\begin{equation}
R_{AD} = \tilde{L}\tilde{V} / \lambda_{AD} = {\cal M}_A \tilde{L} \nu_{ni}/v_A,
\end{equation}
which must fall below unity for ambipolar diffusion to be important,
where $\tilde{L}$ and $\tilde{V}$ are the characteristic length and
velocity scales, and ${\cal M}_A = \tilde{V}/v_A$ is the characteristic
Alfv\'en Mach number. In our situation we again can take the rms velocity as
typical value for $\tilde{V}$.
By setting $R_{AD} = 1$, we can  derive a critical length scale
below which ambipolar diffusion is important
\begin{equation} \label{lcrit1}
\tilde{L}_{cr} = \frac{v_A}{{\cal M}_A \nu_{ni}} 
         \approx (0.041 \mbox{ pc}) \frac{B_{10}}{{\cal M}_A x_6 n_3^{3/2}},
\end{equation}
where the magnetic field strength $B = 10 B_{10}\, \mu G$, the
ionization fraction $x = 10^{-6} x_6$, the neutral number density $n_n
= 10^3 n_3\,{\rm cm}^{-3}$, and we have taken $\rho_n = \mu n_n$, with $\mu =
2.36\,m_{\rm H}$.  

We can attempt to  compare this value to the numerical
dissipation scale by directly computing the ratio of the thermal Jeans length 
$\lambda_J$ that we use to scale our models (as discussed in
\S~\ref{subsec:scaling}) to $\tilde{L}_{cr}$.  We do this
by assuming that  ionization and magnetic field both depend
on the density of the region, following the empirical laws $n_i = 3 \times
10^{-3}\,\mbox{ cm}^{-3}\,(n_n / 10^5\,\mbox{ cm}^{-3})^{1/2}$ (e.g.\
Mouschovias 1991), and $B_{10} \sim 3\,n_3^{1/2}$ (e.g.\ the
observational summary of Crutcher 1999).  We can then find the
interesting result that
\begin{equation}
\frac{\lambda_J}{\tilde{L}_{cr}} = 16.1\,{\cal M}_A 
             \left(\frac{c_{\rm s}}{0.2 \mbox{ km s}^{-1}}\right).
\end{equation}
Crutcher (1999) suggested that typical values of the Alfv\'en
Mach number ${\cal M}_A$ are only slightly above unity. With the
 value in our simulations $\lambda_{\rm J} = 0.1 \sqrt{8 \pi} \approx 0.5$
(cf.~\S3.4) and noting that our cube has a side length of $L =2
\approx 4\lambda_{\rm J}$, this
implies that the critical length scale on which ambipolar diffusion
becomes important in our model units is $\tilde{L}_{cr} =
L/64$.  This is comparable to or even slightly greater than the length
on which numerical dissipation acts in our highest resolution models.
Thus, we can conclude somewhat surprisingly that we may be close to
capturing the full dissipation-free range available to real molecular
clouds in our models.

\section{Fourier Analysis}
\label{sec:wave-mode-analysis}
In this section we discuss the energy distribution on different
spatial scales during various stages of the dynamical evolution of the
system.  We perform a Fourier analysis of the energy, computing the
power spectra of kinetic and potential energies. To allow for a direct
comparison, all models are analyzed on a Cartesian grid with $128^3$
cells. For the SPH models this is done using the kernel smoothing
algorithm, while the $256^3$-ZEUS models are simply degraded in
resolution.  For each cell the potential and kinetic energy content is
calculated, and the kinetic energy is further decomposed into its
solenoidal (rotational) and compressional parts. These quantities are
then all transformed into Fourier space, to find the contribution of
different dimensionless wave numbers $k$, or equivalently, to find the
distribution of energy over different spatial scales $\lambda_k =
L/k$.

\begin{figure}[h]
\unitlength1.0cm
\begin{picture}(16,11.5)
\put(-2.40,-9.00){\epsfbox{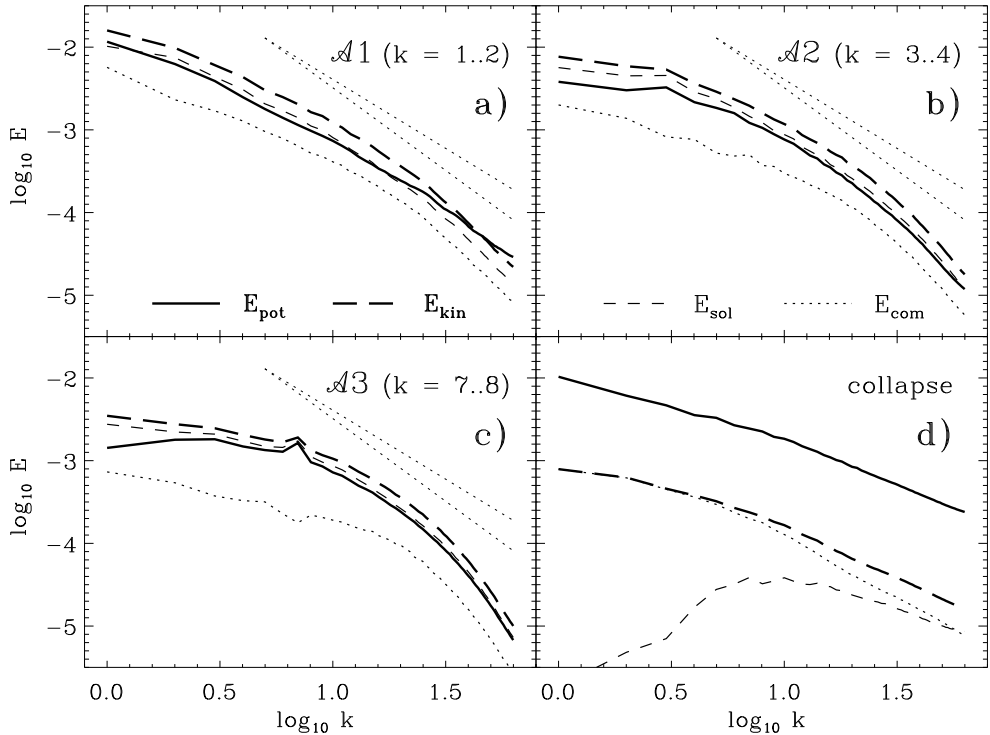}}
\end{picture}
\caption{\label{fig:wave-mode-analysis-1} Energy as function of wave
number $k$ for models with different driving scale: (a) ${\cal A}1$
with $k=1-2$, (b) ${\cal A}2$ with $k=3-4$ and (c) ${\cal A}3$ with
$k=7-8$. The simulations are studied at $t=0.0$, when the hydrodynamic
turbulence is fully developed, immediately after gravity is included.
The plots show potential energy $E_{\rm pot}$ (thick solid lines),
kinetic energy $E_{\rm kin}$ (thick long-dashed lines), its solenoidal
component $E_{\rm sol}$ (short-dashed lines) and its compressional
component $E_{\rm com}$ (dotted lines). The thin dotted lines indicate
the slope $-5/3$ expected from the Kolmogorov (1941) theory and the
slope $-2$ expected for velocity discontinuities associated with
shocks.  For comparison, plot (d) shows a strongly self-gravitating
model that completely lacks turbulent support and therefore contracts
on all scales (data from Klessen et al.~1998).  The energy spectra are
computed on a $128^3$ grid onto which the SPH particle distribution
has been assigned using the kernel smoothing procedure.  }
\end{figure}

The energy spectrum of fully developed turbulence for small-, medium-
and large-scale driving is shown in figure
\ref{fig:wave-mode-analysis-1}. It shows the SPH models (a) ${\cal
A}1$, (b) ${\cal A}2$ and (c) ${\cal A}3$  just at the time $t=0.0$
when gravity is turned on.  In each plot the thick solid lines describe the
potential energy as a function of wave number $k$, and the 
thick long-dashed lines represent the kinetic energy, which can be
decomposed into its solenoidal (rotational) and compressional
components. They are defined via the velocities by $\vec{\nabla}
\cdot \vec{v}_{\rm sol} = 0$ and $\vec{\nabla} \times \vec{v}_{\rm
com} =0$, respectively. 

The models ${\cal A}1$ and ${\cal A}2$, which are driven at long and
intermediate wave lengths ($k=1-2$ and $k=3-4$), appear to exhibit an
inertial range below the driving scale, i.e.\ between $0.5 \sil
\log_{10} k \sil 1.5$.  Note that, in real clouds, the dissipation
scale may lie near the upper end of this wave number range as
discussed in \S~\ref{subsec:real-clouds}.  In this interval the energy
distribution  approximately follows a power law very similar to that predicted by
the Kolmogorov (1941) theory ($E_{\rm kin} \propto k^{-5/3}$). This is
understandable given that, in our models, once turbulence is fully
established, the solenoidal component of the kinetic energy always
dominates over the compressible one, $E_{\rm sol} > E_{\rm com}$.  For
a pure shock dominated flow ($E_{\rm com} \gg E_{\rm sol}$) one would
expect a power spectrum with slope $-2$ (Passot et al.~1988). To guide
the eye, both slopes are indicated as thin dotted lines in plots (a)
to (c). For model ${\cal A}3$ the smaller number of available modes
between the driving scale $k=7-8$ and the Nyquist frequency does not
allow for an unambiguous identification of a turbulent inertial range.
The permanent energy input necessary to sustain an equilibrium state
of turbulence produces a signature in the energy distribution at the
driving wave length. This is most clearly visible in
figure~\ref{fig:wave-mode-analysis-1}c.

The system is globally stable against gravitational collapse, as
indicated by the fact that for every wave number $k$ the kinetic energy
exceeds the potential energy. For comparison we plot in
figure~\ref{fig:wave-mode-analysis-1}d the energy distribution of a
system without turbulent support. The data are taken from Klessen et
al.~(1998) and stem from an SPH simulation with $500\,000$ particles
containing 220 thermal Jeans masses and no turbulent velocity field,
but otherwise identical physical parameters.  The snapshot is taken at
$t=0.2 \tau_{\rm ff}$ after the start of the simulation. 
This system contracts on {\em all} scales and forms
stars at very high rate within a few free-fall times $\tau_{\rm
ff}$. Contrary to the case of hydrodynamic turbulence, the kinetic
energy distribution is dominated by compressional modes, especially at
small wave numbers. The overall energy budget is determined by the
potential energy $E_{\rm pot}$, which outweighs the kinetic energy
$E_{\rm kin}$ on all spatial scales $k$ by about an order of
magnitude.

\begin{figure}[h]
\unitlength1.0cm
\begin{picture}(16,19.50)
\put(-2.40,-4.20){\epsfbox{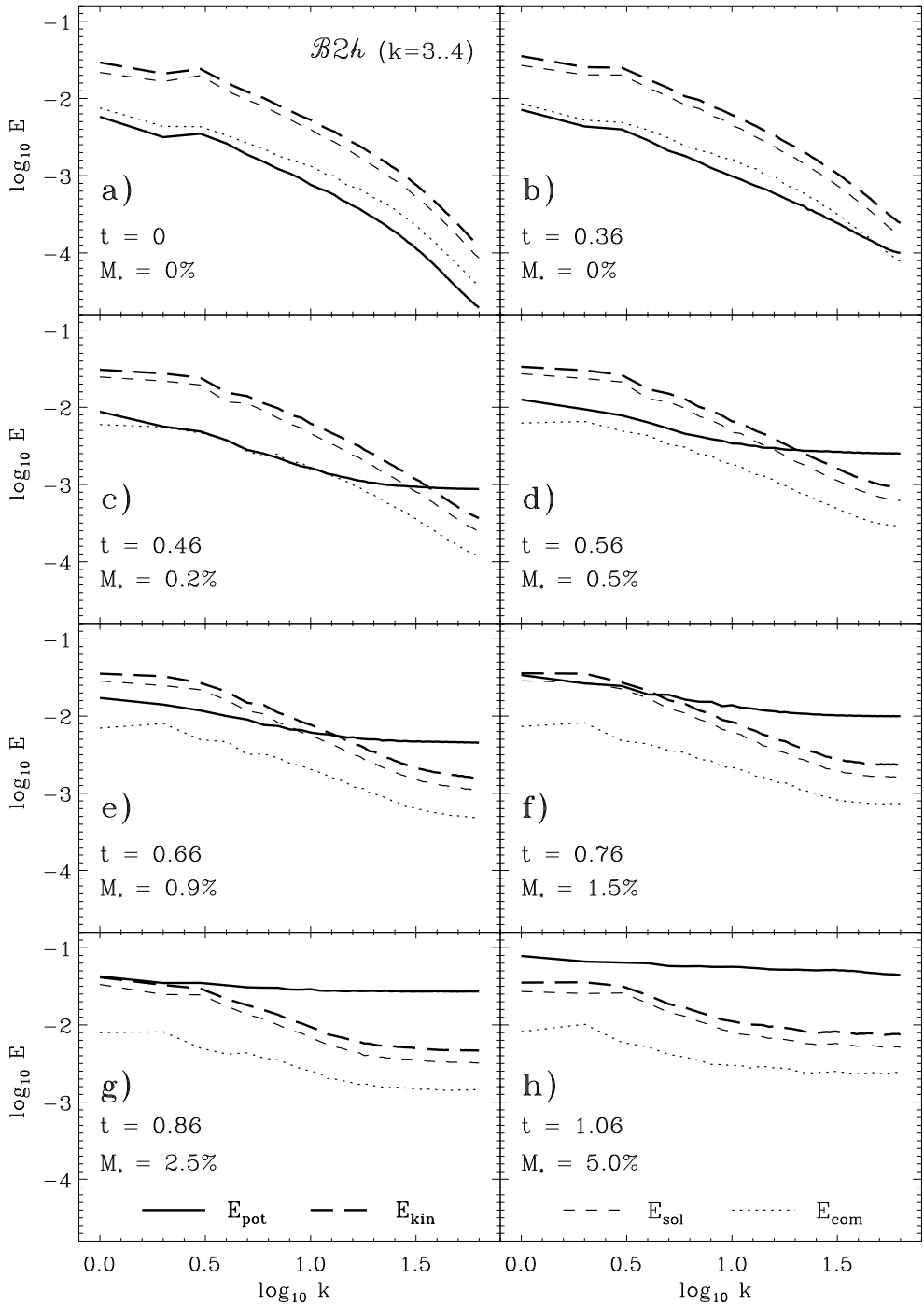}}
\end{picture}
\caption{\label{fig:wave-mode-analysis-2} Fourier analysis of the
high-resolution model ${\cal B}2h$ ($\langle M_{\rm J}\rangle_{\rm
turb} = 3.2$ and $k=3-4$) at different stages of its dynamical
evolution indicated on each plot. Notation and scaling are the same as
in figure \ref{fig:wave-mode-analysis-1}.  Again, the SPH model is
sampled on a $128^3$ mesh.}
\end{figure}

Figure \ref{fig:wave-mode-analysis-2} concentrates on model ${\cal
B}2h$ with $\langle M_{\rm J}\rangle_{\rm turb}=3.2$ and $k=3-4$. It
describes the time evolution of the energy distribution.
Figure~\ref{fig:wave-mode-analysis-2}a shows the state of fully
established turbulence for this model just when gravity is turned on
($t=0.0$).  In the subsequent evolution, local collapse occurs in
shock-generated density enhancements where the potential energy
dominates over the kinetic energy.  This affects the {\em small}
scales first, as seen in the plotted time sequence. As collapse
progresses to higher and higher densities, the scale where the
potential energy dominates rapidly grows. Once the mass fraction in
dense cores has reached about $\sim 3$\%, the potential energy
outweighs the kinetic energy on all scales.  However, this should not
be confused with the signature of global collapse.  The power spectrum
of the potential energy is constant for all $k$. It is the Fourier
transform of a delta function. Local collapse has produced point-like
high-density cores.  The overall filling factor of collapsing clumps
and cores is very low, so most of the volume is dominated by
essentially pure hydrodynamic turbulence. As a consequence, the
velocity field on large scales is not modified much (besides a shift
to higher energies). On small scales, however, the flow is strongly
influenced by the presence of collapsed cores which is noticeable as a
flattening of the power spectra at large wave numbers.  Despite their
small volume filling factor, the cores dominate the overall power
spectrum.  The solenoidal part of the kinetic energy always dominates
over the compressional modes and also the signature of the driving
source in the energy spectrum remains, visible as a `bump' in the
kinetic energy spectrum at $k \approx 8$.

\begin{figure}[h]
\unitlength1.0cm
\begin{picture}(16,20.50)
\put(-2.40,-4.00){\epsfbox{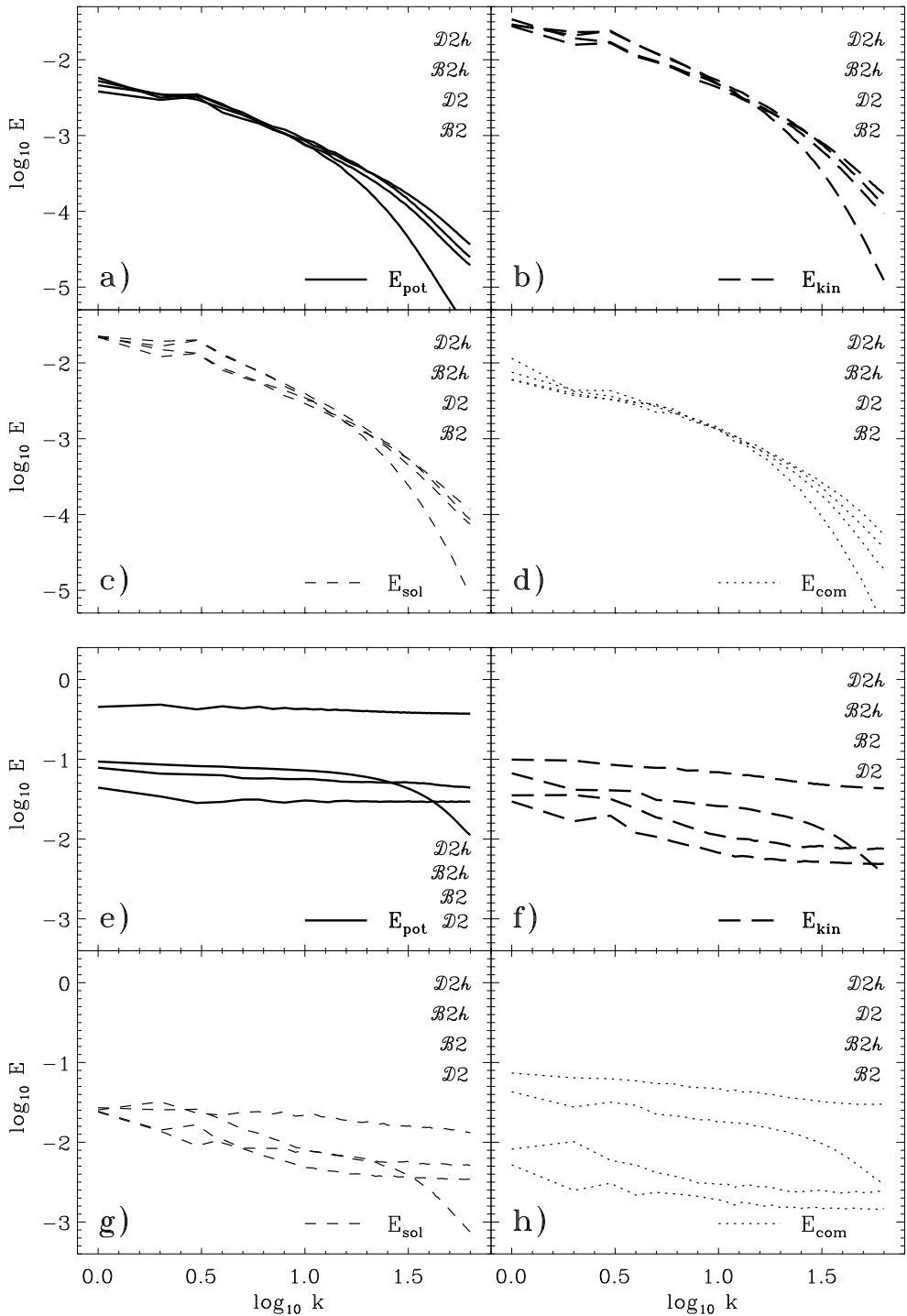}}
\end{picture}
\caption{\label{fig:wave-mode-analysis-3} Wave mode comparison between
four models with identical physical parameters ($\langle M_{\rm
J}\rangle_{\rm turb} = 3.2$ and $k=3-4$) computed with different
numerical methods and resolution: SPH models ${\cal B}2$ and ${\cal
B}2h$ with $50\,000$ and $200\,000$ particles, and ZEUS models ${\cal
D}2$ and ${\cal D}2h$ with $128^3$ and $256^3$ grid cells. To enable
direct comparison, equivalent energy components of all four models are
plotted in each panel. The upper half (a -- d) of the figure shows the
energy distribution of a state of fully developed hydrodynamic
turbulence without gravity. The lower half (e -- h) depicts the system
after gravity is included, when $M_*=5$\% of the total mass is
collapsed onto dense cores. Again, all spectra are computed on a grid
with $128^3$ cells. The labels  refer to the final point of
each spectrum at $k=64$  counted from top to bottom. }
\end{figure}

To show that the {\em global} features of our models are insensitive
to the numerical method used, in figure \ref{fig:wave-mode-analysis-3}
we compare the energy spectra of four different simulations with
identical physical parameters. Analogously to
figure~\ref{fig:2D-cuts}, we chose simulations ${\cal B}2$, ${\cal
B}2h$, ${\cal D}2$ and ${\cal D}2h$ which all have $\langle M_{\rm
J}\rangle_{\rm turb}=3.2$ and $k=3-4$. Models ${\cal B}2$ and ${\cal
B}2h$ are SPH simulations with $50\,000$ and $200\,000$ particles,
while ${\cal D}2$ and ${\cal D}2h$ were calculated using the ZEUS code
with a resolution of $128^3$ and $256^3$ grid zones,
respectively. Figures \ref{fig:wave-mode-analysis-3}a--d directly
compare the different energy components in the four models at $t=0$,
at the stage of fully developed pure hydrodynamic turbulence just
before gravity is switched on. The sequence
\ref{fig:wave-mode-analysis-3}e--h does the same after gravity has
been switched on and the first collapsed cores have formed at
$t_{5\%}$, when the mass accumulated in dense cores is $M_* = 5$\% of
the total mass. This state is identical to the one depicted in figure
\ref{fig:2D-cuts}, allowing for direct comparison.

Comparing the spectra of the different models during the stage of pure
hydrodynamic turbulence (figures \ref{fig:wave-mode-analysis-3}a--d)
shows excellent agreement between the energy spectra of the different
models, suggesting the energy distribution is well converged.  Between
the scales of energy input (at $k=3-4$) and diffusive energy loss, all
the spectra follow the same power law with slope $-5/3$ (analogous to
the spectra shown in figure \ref{fig:wave-mode-analysis-1}).  The
dissipation scale manifests itself as a drop-off from the power-law at
large wave numbers.  The inertial range of turbulence is largest in
the high-resolution ZEUS model ${\cal D}2h$, where it spans about one
order of magnitude in $k$.  The high-resolution SPH and
medium-resolution ZEUS models ${\cal B}2h$ and ${\cal D}2$ have
inertial ranges nearly as long.  The medium-resolution SPH model
${\cal B}2$ has the shortest range with $\Delta \log_{10} k \approx
0.5$.  At wave numbers above the dissipation scale, our results are
converged to better than 10\% in the log of energy.

In the presence of self-gravity, the energy spectra are no longer well
converged.  The actual density contrast reachable in collapsing cores,
or to a lesser extent in shock fronts, depends on the numerical
resolution and algorithm used (see \S~\ref{sec:method}).  The same
applies to the potential energy and to the compressional component of
the kinetic energy. Figures \ref{fig:wave-mode-analysis-3}e and h
therefore exhibit significant differences between the various models.
These differences are much smaller for the solenoidal part of the
kinetic energy, which measures rotational motions and is therefore
less sensitive to strong density contrasts in small volumes.
Variations in the total kinetic energy distribution are mainly due to
differences in the compressional modes.

The rapid energy decrease for wave numbers $\log_{10} k >1.4$
in the grid-based model ${\cal D}2$ is due to the fact that these
scales approach the grid resolution.  A similar decrease would be seen 
in the other three models if they were sampled at wave numbers all the way
up to the effective resolution (grid size for ZEUS or smoothing length for SPH).
Remember that all spectra shown here are
computed on the same grid with a linear resolution of 128
cells. Despite the fact that the $256^3$ ZEUS model ${\cal D}2h$ has
been resampled and degraded in resolution, large density contrasts 
still occur on the smallest scales of the resampled grid. The energy
spectra therefore remain flat towards the Nyquist wave number. Similarly, the use
of adaptive particle smoothing lengths in SPH  allows the resolution of
dense cores smaller than the cell size of the
$128^3$ grid used for the energy sampling. Again, there is no
loss of power towards the Nyquist wave number of the spectra.
However, high resolution in high density
regions is achieved at the cost of low resolution in voids. As low
density regions occupy most of the volume, on large scales the SPH simulations
tend 
to have lower energy content than the grid-based models.  

Understanding these various
effects allows us to understand what questions we can reasonably ask of
these simulations.  The presence or absence of collapse and the distribution of
kinetic energy on large scales are questions for 
which we can give well-converged answers, but the details of the strength of
that collapse still depend on the details of the numerical method and should
not be used quantitatively.

\section{Implications for Star Formation in Molecular Clouds}
\label{sec:SF-in-MC}
In \S~\ref{sec:collapse} we have shown that the rate and efficiency of
local collapse in turbulent molecular clouds depend on the strength
and the effective wave length of the driving energy input.  Star
formation will follow local collapse (e.g.~V{\'a}zquez-Semadeni,
Canto, \& Lizano 1998), so we can use these properties of our
turbulence models to try to explain the observed spatial and age
distributions of young stars in molecular clouds.  We use the spatial
and age distributions of sink particles generated in the SPH models
with different parameters for this purpose.

\subsection{Clustered vs.\ Isolated Star Formation}
\label{subsec:clustered-isolated-SF}

Different star formation regions present different distributions of
protostars and pre-main sequence stars.  In some regions, such as the
Taurus molecular cloud, stars form isolated from other stars,
scattered throughout the cloud (Mizuno et al.~1995).  In other regions, they form in
clusters, as in L1630 in Orion (Lada 1992), or even more extremely in
starburst regions such as 30 Doradus (Walborn et al.\ 1999). 

>From the simulations presented here, it is evident that the {\em
length scale} and {\em strength} at which energy is inserted into the
system determine the structure of the turbulent flow and therefore the
locations at which stars are most likely to form. Large-scale driving
leads to large coherent shock structures (see
e.g.~figure~\ref{fig:3D-cubes-1..2+7..8}a). Local collapse occurs
predominantly in filaments and layers of shocked gas and is very
efficient in converting gas into stars. This leads to what we can
identify as `clustered' mode of star formation: stars form in coherent
aggregates and clusters. Even more so, this applies to regions of
molecular gas that have become decoupled from energy input. As
turbulence decays, these regions begin to contract and form dense
clusters of stars with very high efficiency on about a free-fall time
scale (Klessen et al.~1998, Klessen \& Burkert 2000).  The same holds
for insufficient support, i.e.~for regions where energy input is not
strong enough to completely balance gravity. They too will contract to
form dense stellar clusters.

The `isolated' mode of star formation occurs in molecular cloud regions that
are supported by driving sources that act on {\em small} scales and in an
incoherent or stochastic manner. In this case, individual shock induced
density fluctuations form at random locations and evolve more or less
independently of each other. The resulting stellar population is widely
dispersed throughout the cloud and, as collapsing clumps are 
exposed to frequent shock interaction, the overall star formation rate
is low.

\begin{figure}[h]
\unitlength1.0cm
\begin{picture}(16,18.50)
\put(-3.80,-10.50){\epsfbox{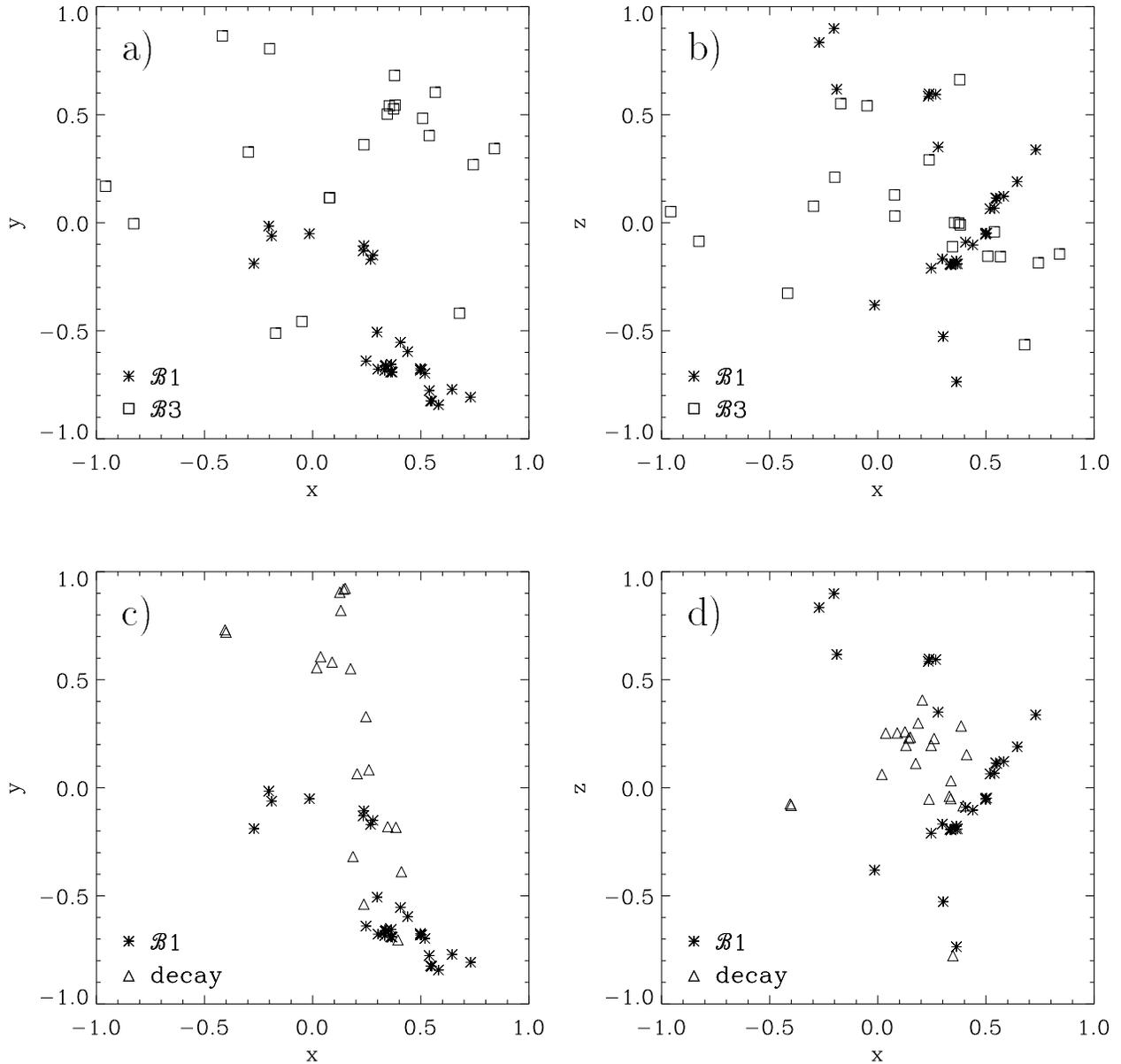}}
\end{picture}
\caption{\label{fig:2D-projection} Comparison of core locations
between two globally stable models with different driving wave length
(${\cal B}1$ with $k=1-2$ and ${\cal B}3$ with $k=7-8$) projected into
(a) the $xy$-plane and into (b) the $xz$-plane. Plots (c) and (d) show
the core locations for model ${\cal B}1$ now contrasted with a
simulation of decaying turbulence from Klessen~(1999).  The snapshots
are selected such that the mass accumulated in dense cores is $M_*
\sil 20$\%. Note the different times needed for the different models
to reach this point.  For  model ${\cal B}1$ data are taken at $t=1.1$,
for ${\cal B}3$ at $t=12.3$. The simulation of freely decaying
turbulence is shown at $t=1.1$. All times are normalized to the
global free-fall time scale of the system.  }
\end{figure}

To demonstrate these points, we compare in
figure~\ref{fig:2D-projection} the distribution of sink particles for
several different models, projected onto the $xy$- and $xz$-planes.
As an example of coherent local collapse, we choose model ${\cal B}1$,
where the turbulence is driven strongly at long wave lengths. The flow
is dominated by large coherent shocks, so cores form in aggregates
associated with the filamentary structure of shock compressed gas
(cf.~with figure~\ref{fig:3D-cubes-1..2+7..8}). The overall efficiency
of converting gas into stars in this `clustered' mode is very high.
The upper half of figure~\ref{fig:2D-projection} compares the model
${\cal B}1$ with model ${\cal B}3$, which is driven at small scales
and results in incoherent collapse behavior. Individual cores form
independently of each other at random locations and random times. In
this `isolated' mode, cores are widely distributed throughout the
entire volume and exhibit considerable age spread.

In the lower half of figure~\ref{fig:2D-projection} we contrast the
large-scale driving model ${\cal B}1$ with a simulation of freely
decaying turbulence described by Klessen (2000) that has the same
thermal Jeans mass.  In decaying turbulence, once the kinetic energy
level has decreased sufficiently, all spatial modes of the system
contract gravitationally, including the global ones. As in the case of
large-scale shock compression, stars form more or less coevally in a
very limited volume with high efficiency. Both insufficient turbulent
support and the complete loss of it therefore appear to lead to
clustered star formation.  The Trapezium cluster in Orion may be a
good example for the outcome of this mechanism (e.g.~Hillenbrand 1997,
Hillenbrand \& Hartmann 1998).  All the projections shown in
figure~\ref{fig:2D-projection} are taken at a stage of the dynamical
evolution when the mass accumulated in dense cores is $M_* \approx
20$\%.  This occurs at very different times, as noted in the captions,
which directly reflects the varying efficiencies of local collapse
in these models.

\begin{figure}[h]
\unitlength1.0cm
\begin{picture}(16,2.00)
\put(-3.80,-10.20){\epsfbox{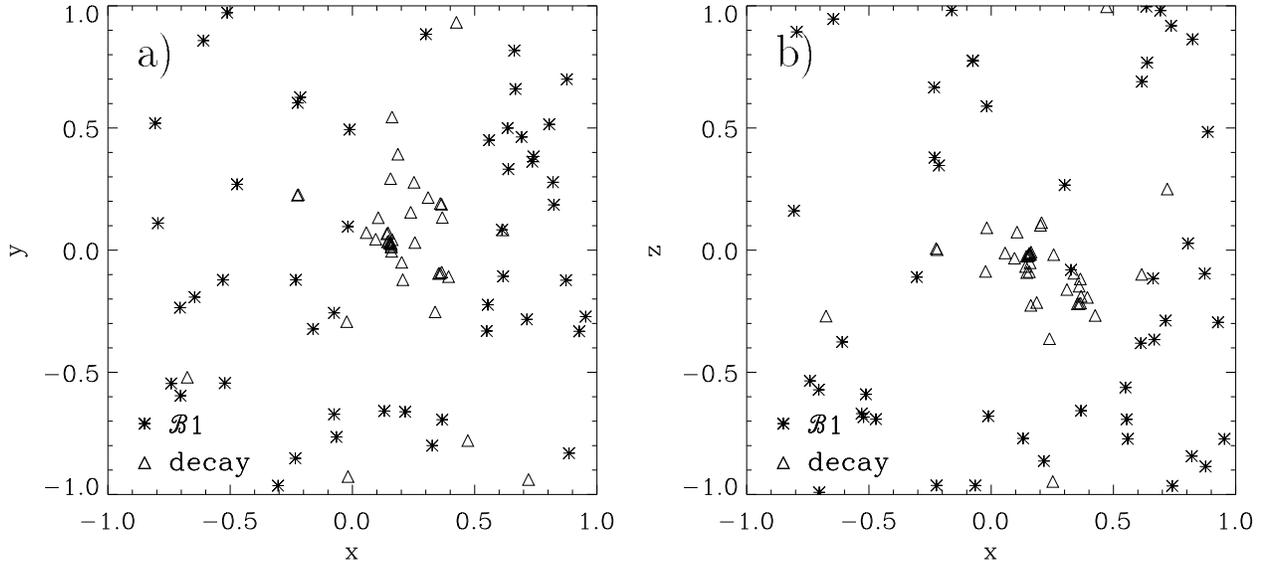}}
\end{picture}
\caption{\label{fig:2D-projection-later-times} Core positions for
model ${\cal B}1$ ($k=1-2$) and the decay model when the core mass
fraction is $M_* \approx 65$\%, projected into (a) the $xy$-plane and
(b) the $xz$-plane (cf.\ figure~\ref{fig:2D-projection}c \& d).  For
${\cal B}1$ the time is $t=8.7$ and for decay model $t=2.1$. Whereas
the cluster in ${\cal B}1$ is completely dissolved and the stars are
widely dispersed throughout the computational volume, the core cluster
in the decay simulation remains bound.}
\end{figure}

 Despite the fact that turbulence that is driven on large scales as
well as turbulence that is freely decaying lead to star formation in
aggregates and clusters, figure~\ref{fig:2D-projection-later-times}
suggests a possible way to distinguish between them by taking the
long-term evolution of the resulting clusters into account. Whereas
decaying turbulence typically leads to the formation of a bound
stellar cluster, the dynamical relaxation of aggregates associated
with large-scale coherent shock fronts quite likely results in their
complete dispersal. This is illustrated in
figure~\ref{fig:2D-projection-later-times}, which compares the core
distribution in model ${\cal B}1$ and in the decay simulation at $M_*
\approx 65$\%, when both systems have already undergone considerable
evolution.  The cores in model ${\cal B}1$ are completely dispersed
throughout the molecular cloud volume. The cluster that formed during
the turbulent decay remains bound with a much longer evaporation time
scale. Note, however, that at late stages of the dynamical evolution
our isothermal model becomes less appropriate as the feedback effects
from newly formed stars are not taken into account. Ionization and
outflows from these stars will likely retard or even prevent the
accretion of the remaining gas onto the protostars. This limits the
validity of our models at very late times.

\subsection{The Time Scales of Star Formation}
\label{subsec:timing-SF}

\begin{figure}[h]
\unitlength1.0cm
\begin{picture}(16,11.50)
\put(-0.5,0){\epsfbox{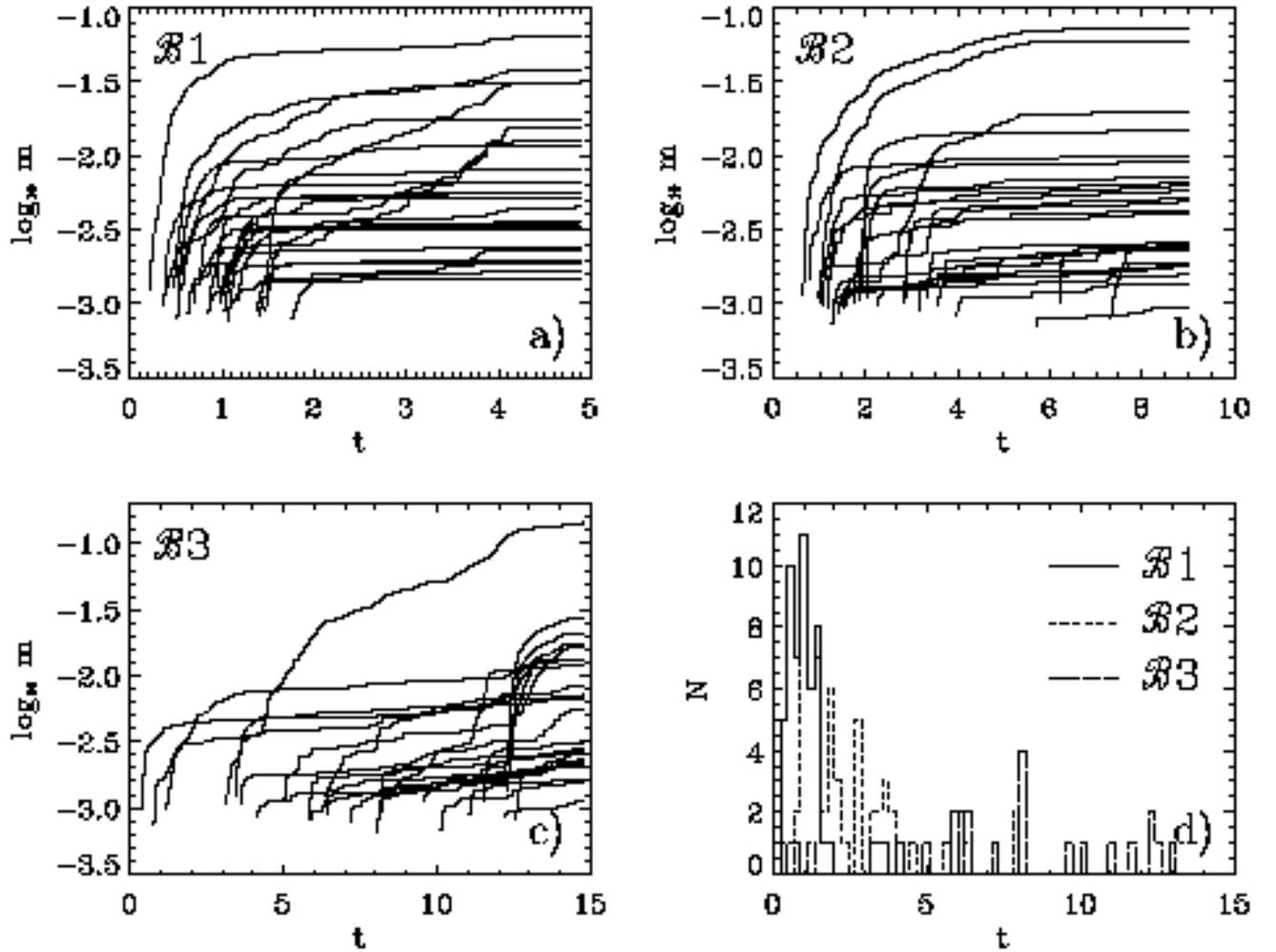}}
\end{picture}
\caption{\label{fig:core-formation-histogram} Core masses as function
of time in SPH models (a) ${\cal B}1$ with $k=1-2$ driving, (b) ${\cal
B}2$ with $k=3-4$ driving, and (c) ${\cal B}3$ with $k=7-8$ driving.
The curves represent the formation and accretion histories of
individual cores. For the sake of clarity, only every other core is
shown in (a) and (b), whereas in (c) the evolution of every single
core is plotted.  Time is given in units of the global free-fall time
$\tau_{\rm ff}$. Note the different time scale in each plot. In the
depicted time interval models ${\cal B}1$ and ${\cal B}2$ reach a core
mass fraction $M_* =70$\%, and both form roughly 50 cores. Model
${\cal B}3$ reaches $M_* =35$\% and forms only 25 cores.  Figure (d)
compares the distributions of formation times. The age spread
increases with decreasing driving scale showing that clustered core
formation should lead to a coeval stellar population, whereas a
distributed stellar population should exhibit considerable age spread.
}
\end{figure}

In the previous section we have argued that stellar clusters form
predominantly in molecular cloud regions that are insufficiently
supported by turbulence or where only large-scale driving is active.
In the absence of driving, molecular cloud turbulence decays more
quickly than the free-fall time scale (Mac Low 1999).  The free-fall
time $\tau_{\rm ff}$ is thus the typical time scale on which dense
stellar clusters will form in the absence of support (Klessen et
al.~1998) or in the presence of decaying turbulence (Klessen 2000).
Even in the presence of support from large-scale driving, collapse
will occur on roughly this time scale, as shown for model ${\cal B}1$
in figure~\ref{fig:core-formation-histogram}a.  If we assume that once
we have identified a dense core it continues to collapse on a very
short time scale to build up a stellar object in its center, then this
spread relates directly to the star formation time scale.  Therefore
the age distribution will be roughly $\tau_{\rm ff}$ for stellar
clusters that form coherently with high star formation efficiency.
When scaled to low densities ($n({\rm H}_2) \approx 10^2\,{\rm
cm}^{-3}$ and $T\approx10\,$K) the global free-fall time scale in our
models is $\tau_{\rm ff} = 3.3 \times 10^6\,$years.  
If star forming clouds such as Taurus indeed have ages of order
$\tau_{\rm ff}$, as suggested by Ballesteros-Paredes et al.\ (1999),
then the long star formation time scales computed here is quite
consistent with the very low star formation efficiencies seen in
Taurus (e.g.\ Leisawitz et al.\ 1989), as the cloud simply has not had
time to form many stars.    In the case of high-density regions
($n({\rm H}_2) \approx 10^5\,{\rm cm}^{-3}$ and $T\approx10\,$K) the
dynamical evolution proceeds much faster and the corresponding
free-fall time scale drops to $\tau_{\rm ff} = 1.0 \times
10^5\,$years.  These values indeed agree well with observational data,
e.g.~the formation time scale of the Orion Trapezium cluster, which is
inferred to stem from gas of density $n({\rm H}_2) \sil 10^5\,{\rm
cm}^{-3}$, is estimated to be less than $10^6$ years (Hillenbrand \&
Hartmann 1998).

The age spread increases with increasing driving wave number $k$ and
increasing $\langle M_{\rm J} \rangle _{\rm turb}$.  Molecular cloud
regions supported against global collapse by driving sources that act
on small scales host stochastical star formation on much longer time
scales and with much lower efficiency.  The process is incoherent and
the expected stellar age spread therefore larger.  Indeed, in
figure~\ref{fig:core-formation-histogram}b, which shows the accretion
history of selected cores in model ${\cal B}2$ with $k=3-4$
(representing more isolated star formation), core formation extends
over a longer period.  This is even more pronounced in model ${\cal
B}3$ with $k=7-8$ shown in
figure~\ref{fig:core-formation-histogram}c. Note that the real time
spread in this model is even larger than suggested by the figure,
because by the time we stopped the simulation the accreted mass
fraction was only $M_*= 35$\%. We expect that more cores would form in
the subsequent evolution. Models ${\cal B}1$ and ${\cal B}2$, on the
other hand, already reach $M_* \approx 70$\% in the time interval
shown. They each form roughly 50 cores, twice as much as model ${\cal
B}3$. For a direct comparison, figure
\ref{fig:core-formation-histogram}d plots the distribution of core
formation times in each of the three models on the same scale.
These long periods of core formation for globally supported clouds 
appear consistent with the low efficiencies of star-formation in regions of 
isolated star formation, such as Taurus, even if they are rather young objects 
with ages of order $\tau_{\rm ff}$.

\section{Summary and Conclusions}
\label{sec:conclusions}

We have studied the conditions that allow self gravity to cause
collapse in a region of supersonic turbulence.  We used this study to
determine whether interstellar turbulence can support molecular clouds
against gravitational collapse, revealing the scales and physical
conditions that allow star formation to occur.  To perform these
studies, we computed numerical simulations of the time evolution of
turbulent, self-gravitating, isothermal gas with two different
computational schemes: a particle-based, Lagrangian method (SPH); and
a second-order, Eulerian, grid-based method (ZEUS).  By comparing
results from these two  different numerical schemes we benefit
from the advantages of both methods, and we are furthermore able to
estimate the influence of algorithm as well as resolution on our
results.  We next summarize and discuss our results.



\begin{enumerate}

\item Supersonic turbulence strong enough to globally support a
molecular cloud against collapse will usually cause {\em local}
collapse.  The turbulence establishes a complex network of interacting
shocks.  The local density enhancements in fluctuations created by
converging shock flows can be large enough to become gravitationally
unstable and collapse. This occurs if the local Jeans length becomes
smaller than the size of the fluctuation.  The probability for this to
happen, the efficiency of the process, and the rate of continuing
accretion onto collapsed cores are strongly dependent on the driving
wave length and on the rms velocity of the turbulent flow, and thus on
the driving luminosity.  Collapse criteria derived from
incompressible, self-gravitating turbulence (Chandrasekhar 1951,
Bonazzola et al.~1987, 1992, V{\'a}zquez-Semadeni \& Gazol 1995)
indeed determine the {\em global} or large-scale collapse properties
of the medium.  However, the occurrence and ubiquity of {\em local}
collapse in shock-generated fluctuations drastically limit the
application of these criteria to interpreting the actual behavior of
star-forming regions, as {\em localized} collapse can still occur even
if the cloud as a whole is stabilized by turbulence.

\item Fluctuations in turbulent velocity fields are highly {\em
transient}.  The random flow that creates local density enhancements
can also disperse them.  For local collapse to result in the formation
of stars, locally Jeans unstable, shock-generated, density
fluctuations must collapse to sufficiently high densities on time
scales shorter than the typical time interval between two successive
shock passages.  Only then are they able to `decouple' from the
ambient flow pattern and survive subsequent shock interactions.  (If
they begin collapse magnetically supercritical, they will remain so
for the rest of the collapse.)  The shorter the time between shock
passages, the less likely these fluctuations are to survive.  Hence,
keeping the scale of energy input fixed and increasing the driving
luminosity leads to a decrease of the star formation efficiency.
Local collapse takes longer to occur and the mass accretion rate onto
cores is reduced.  Similarly, driving on small scales leads to a lower
star formation rate than driving on larger scales at the same rms
velocity.  Quantitatively, our models appear to show that it is
possible to prevent 95\% of the gas from collapsing into dense cores
over ten global free-fall times with strong enough driving on short
enough wave lengths.  If a physical mechanism for such driving can be
found, this could indeed explain the long cloud life times and low
star formation rates commonly ascribed to Galactic molecular clouds
(Blitz \& Shu 1980, Blitz 1993).  Conversely, if such driving does not
exist, then molecular clouds should be transient objects and the short
life times proposed by Ballesteros-Paredes et al. (1999) and Elmegreen
(2000) appear more likely.

\item Local collapse can only be halted completely if the turbulent
driving mechanism supplies enough energy on scales {\em smaller} than
the Jeans length of the `typical' fluctuation. In supersonic
turbulence the typical density contrast is $\delta \rho \propto {\cal
M}^2$, where ${\cal M}$ is the {\em rms} Mach number of the
flow.  Thus, the Jeans length is reduced by a factor of ${\cal M}$ 
with respect to the global value.  Complete prevention of local
collapse requires even stronger and shorter wave length driving, as
there will be stochastic turbulent fluctuations with even larger
density contrast. However, the time scale for the occurence of high density
fluctuations increases rapidly with $\delta \rho$, so sufficiently
strong driving can prevent local collapse for arbitrarily long periods
of time.  Such strong driving may be rather difficult to arrange in a
real molecular cloud, however. 

If we assume that stellar driving sources have an effective wave
length close to their separation, then the condition that driving acts
on scales smaller then the Jeans wave length in `typical' shock
generated gas clumps requires the presence of an extraordinarily large
number of stars evenly distributed throughout the cloud, with typical
separation 0.1 pc in Taurus, or only 350 AU in Orion (taking our fully
supported case as an example).  This is not observed.  Mac Low et al.\
(1999) show that magnetic fields probably cannot transfer energy
efficiently enough to small scales either.  Furthermore, ambipolar
diffusion may begin to damp turbulent motions at these scales.  Unless
some other mechanism can force energy onto these small scales, local
collapse will occur within globally supported molecular clouds.  In
addition, very small driving scales seem to be at odds with the
observed velocity fields at least in some molecular clouds (e.g. Heyer et
al.~1997 for the Cep OB3 cloud).

\item Interstellar clouds driven on large scales or without even
global turbulent support very rapidly form stars in clusters.  Gas
collapses into dense cores  within a few free-fall times and the star
formation efficiency is more than 50\%.  On the contrary, in gas that
is supported by turbulence driven at small scales, local collapse
occurs sporadically over a large time interval, forming isolated
stars.  The total star formation efficiency before the cloud dissolves
due to stellar feedback or external shocks will probably be low.  Thus, the
strength and 
nature of the turbulence may be fully sufficient to explain the
difference between the observed isolated and clustered modes of star
formation.  

\item In turbulent flows, it is impossible to predict from the start
when and where individual cores form and how they evolve. Firm
statistical results can, however, be derived from analyzing large
ensembles of cores and from characterizing other global indicators of
the dynamical state of the system such as total potential and kinetic
energy.  In all our models except the ones driven below the
fluctuation Jeans scale, gravity eventually begins to dominate over
kinetic energy. This first occurs on small scales, indicating the
presence of local collapse. As dense collapsed cores form, the power
spectrum of the gravitational energy becomes essentially flat. The
kinetic energy, on the other hand, appears to follow at intermediate
wave numbers a Kolmogorov power spectrum with slope $-5/3$, less steep
than the spectrum expected for pure shock flows. The slope remains
almost unaltered during the course of the evolution, indicating that a
large volume fraction of the system is always well described by pure
hydrodynamic turbulence.  The spatial extent of collapsing regions
(where infall motions dominate over the turbulent flow) is relatively
small. This also explains the fact that the solenoidal component of
the flow always dominates over the compressional part.

\end{enumerate}

\acknowledgments We thank A.~Burkert, E.~V\'azquez-Semadeni, and
E.~Zweibel for valuable discussions, and P.~Padoan for an insightful
and exceptionally prompt referee's report.  Computations presented
here were performed on GRAPE processors at the MPI for Astronomy and
at the Sterrewacht Leiden, and on SGI Origin 2000 machines of the
Rechenzentrum Garching of the Max-Planck-Gesellschaft, the National
Center for Supercomputing Applications (NCSA), and the Hayden
Planetarium.  ZEUS was used by courtesy of the Laboratory for
Computational Astrophysics at the NCSA. This research has made use of
NASA's Astrophysics Data System Abstract Service.

\newpage
\begin{table}
\begin{center}
\begin{tabular}{cccccccc}
\hline
Name &  Method  & Resolution    & $k_{\rm drv}$ & $\dot{E}_{\rm in}$ & 
$E_{\rm kin}^{eq}$ & $\langle M_{\rm J}\rangle_{\rm turb}$ & $t_{5\%}$  \\
\hline
${\cal A}1$  &  SPH  & $200\,000$ & $1-2$ & $0.1$ & $0.15$ & $0.6$  & $0.5$  \\
${\cal A}2$  &  SPH  & $200\,000$ & $3-4$ & $0.2$ & $0.15$ & $0.6$  & $0.7$  \\
${\cal A}3$  &  SPH  & $200\,000$ & $7-8$ & $0.4$ & $0.15$ & $0.6$  & $2.2$  \\
\hline                                                                     
${\cal B}1$  &  SPH  & $50\,000$ &   $1-2$ & $0.5$ &  $0.5$ & $3.2$  &  $0.5$ \\
~$\,{\cal B}1h$ &  SPH  & $200\,000$&$1-2$ & $0.5$ &  $0.5$ & $3.2$  &  $0.4$ \\
~${\cal B}2\ell$ &  SPH  & $20\,000$&$3-4$ & $1.0$ &  $0.5$ & $3.2$  &  $1.6$ \\
${\cal B}2$  &  SPH  & $50\,000$ &   $3-4$ & $1.0$ &  $0.5$ & $3.2$  &  $1.5$ \\
~$\,{\cal B}2h$ &  SPH  & $200\,000$&$3-4$ & $1.0$ &  $0.5$ & $3.2$  &  $1.4$ \\
${\cal B}3$  &  SPH  & $50\,000$ &   $7-8$ & $2.4$ &  $0.5$ & $3.2$  &  $6.0$ \\
${\cal B}4$  &  SPH  & $50\,000$ & $15-16$ & $5.0$ &  $0.5$ & $3.2$  & $8.0$  \\
${\cal B}5$  &  SPH  & $50\,000$ & $[39-40]$ & $[5.9]$ &  $[0.3]$ & $[1.7]$  &  ---  \\ 
\hline                                                                     
${\cal C}2$  &  SPH  & $50\,000$ &   $3-4$ & $7.5$ &  $2.0$ & $18.2$  &  $6.0$ \\
\hline                                                                     
~${\cal D}1\ell$ & ZEUS  &  $64^3$ & $1-2$   & $0.4$ &  $0.5$ & $3.2$  &  *   \\
~${\cal D}2\ell$ & ZEUS  &  $64^3$ & $3-4$   & $0.8$ &  $0.5$ & $3.2$  &  *   \\
~${\cal D}3\ell$ & ZEUS  &  $64^3$ & $7-8$   & $1.6$ &  $0.5$ & $3.2$  &  *   \\
${\cal D}1$      & ZEUS  & $128^3$ & $1-2$   & $0.4$ &  $0.5$ & $3.2$  &$0.4$ \\
${\cal D}2$      & ZEUS  & $128^3$ & $3-4$   & $0.8$ &  $0.5$ & $3.2$  &$1.2$ \\
${\cal D}3$      & ZEUS  & $128^3$ & $7-8$   & $1.6$ &  $0.5$ & $3.2$  &$2.4$ \\
${\cal D}5$      & ZEUS  & $128^3$ & $[39-40]$ &[8.3] &  $[0.5]$ & $[3.2]$  & ---  \\
~${\cal D}1h$    & ZEUS  & $256^3$ & $1-2$   & $0.4$ &  $0.5$ & $3.2$  &$0.4$ \\
~${\cal D}2h$    & ZEUS  & $256^3$ & $3-4$   & $0.8$ &  $0.5$ & $3.2$  &$1.2$ \\
~${\cal D}3h$    & ZEUS  & $256^3$ & $7-8$   & $1.6$ &  $0.5$ & $3.2$  &$3.1$ \\
\hline
\end{tabular}
\end{center}
\caption{\label{tab:models}
Overview of the models.  The columns give model name,
numerical method, resolution, driving wave lengths $k$, energy input
rate $\dot{E}_{\rm in}$, equilibrium value of kinetic energy without
self-gravity $E_{\rm kin}^{eq}$, turbulent Jeans mass $\langle M_{\rm
J}\rangle_{\rm turb}$, and the time required to reach a core mass
fraction $M_* = 5$\%.  The resolution is given for SPH as particle
number and for ZEUS as number of grid cells. Dashes in the last column
indicate that no sign of local collapse was observed within
$20\tau_{\rm ff}$, while stars indicate that the numerical resolution
was insufficient for unambiguous identification of collapsed
cores. The total mass in the system is $M=1$. Models ${\cal B}5$ and
${\cal D}5$ focus on a subvolume with mass $M=0.25$ and
decreased sound speed $c_{\rm s} = 0.05$. They are driven with $k=9-10$
and $\dot{E}_{\rm in}= 0.06$. When scaled up to the standard cube this
corresponds to the {\em effective} values given in square brackets.
Model ${\cal B}2$ has been calculated five times with different random
initializations.  The additional models are not listed separately, but are
called ${\cal B}2^{a}$ -- ${\cal
B}2^{d}$ in the text.}
\end{table}


%

\newpage

\end{document}